\documentclass[pdflatex,sn-mathphys-num]{sn-jnl}

\usepackage{graphicx}%
\usepackage{multirow}%
\usepackage{amsmath,amssymb,amsfonts}%
\usepackage{amsthm}%
\usepackage{mathrsfs}%
\usepackage[title]{appendix}%
\usepackage{xcolor}%
\usepackage{textcomp}%
\usepackage{manyfoot}%
\usepackage{booktabs}%
\usepackage{algorithm}%
\usepackage{algorithmicx}%
\usepackage{algpseudocode}%
\usepackage{listings}%
\usepackage{threeparttable}
\usepackage{float}
\usepackage{subcaption}
\usepackage{placeins}
\usepackage{ragged2e} % for better wrapping
\usepackage{tabularx}
\usepackage{pifont} % for \ding{55} (✗ symbol)

\colorlet{punct}{red!60!black}
\definecolor{background}{HTML}{EEEEEE}
\definecolor{delim}{RGB}{20,105,176}
\colorlet{numb}{magenta!60!black}

\lstdefinelanguage{json}{
    basicstyle=\normalfont\ttfamily,
    numbers=left,
    numberstyle=\scriptsize,
    stepnumber=1,
    numbersep=8pt,
    showstringspaces=false,
    breaklines=true,
    frame=lines,
    backgroundcolor=\color{background},
    literate=
     *{0}{{{\color{numb}0}}}{1}
      {1}{{{\color{numb}1}}}{1}
      {2}{{{\color{numb}2}}}{1}
      {3}{{{\color{numb}3}}}{1}
      {4}{{{\color{numb}4}}}{1}
      {5}{{{\color{numb}5}}}{1}
      {6}{{{\color{numb}6}}}{1}
      {7}{{{\color{numb}7}}}{1}
      {8}{{{\color{numb}8}}}{1}
      {9}{{{\color{numb}9}}}{1}
      {:}{{{\color{punct}{:}}}}{1}
      {,}{{{\color{punct}{,}}}}{1}
      {\{}{{{\color{delim}{\{}}}}{1}
      {\}}{{{\color{delim}{\}}}}}{1}
      {[}{{{\color{delim}{[}}}}{1}
      {]}{{{\color{delim}{]}}}}{1},
}
\hypersetup{
    colorlinks=true,
    linkcolor=blue,
    filecolor=magenta,      
    urlcolor=cyan,
    pdftitle={Overleaf Example},
    pdfpagemode=FullScreen,
}

\def\tsc#1{\csdef{#1}{\textsc{\lowercase{#1}}\xspace}}
\tsc{WGM}
\tsc{QE}
\tsc{EP}
\tsc{PMS}
\tsc{BEC}
\tsc{DE}

\theoremstyle{thmstyleone}%
\theoremstyle{thmstyletwo}%

\theoremstyle{thmstylethree}%

\begin{document}

\title[Article Title]{HySecTwin: A Knowledge-Driven Digital Twin Framework Augmented with Hybrid Reasoning for Cyber-Physical Systems}


\author[1,2]{\fnm{David} \sur{Holmes}}\email{d.holmes@ecu.edu.au}

\author*[2,3]{\fnm{Ahmad} \sur{Moshin}}\email{a.moshin@ecu.edu.au}
\equalcont{These authors contributed equally to this work.}

\author[1,2]{\fnm{Surya} \sur{Nepal}}\email{surya.nepal@data61.csiro.au}
\equalcont{These authors contributed equally to this work.}

\author[1,2]{\fnm{Leslie} \sur{Sikos}}\email{l.sikos@ecu.edu.au}
\equalcont{These authors contributed equally to this work.}

\author[1,2]{\fnm{Iqbal} \sur{Sarker}}\email{m.sarker@ecu.edu.au}
\equalcont{These authors contributed equally to this work.}

\author[1,2]{\fnm{Helge} \sur{Janicke}}\email{h.janicke@ecu.edu.au}
\equalcont{These authors contributed equally to this work.}

\affil*[1]{\orgname{Edith Cowan University}, \orgaddress{\street{270 Joondalup Drive}, \city{Joondalup}, \postcode{6027}, \state{Western Australia}, \country{Australia}}}

\affil[2]{\orgname{CSIRO}, \orgdiv{Data61}, \city{Sydney}, \state{NSW}, \country{Australia}}

%%==================================%%
% Abstract %%
%%==================================%%

\abstract
{Existing Digital Twin (DT) approaches lack semantic reasoning capabilities for effective cybersecurity modelling in Cyber-Physical Systems (CPS). This paper presents HySecTwin, a knowledge-driven digital twin architecture that places automated reasoning at the core of real-time threat detection. HySecTwin incorporates semantic modelling to transform heterogeneous CPS telemetry, device attributes, and operational relationships into machine-interpretable representations, combined with an embedded reasoning engine operating over contextualized system states. Unlike opaque detection methods, the framework integrates deterministic rule-based inference with hybrid fuzzy reasoning to generate explicit, interpretable, and auditable security assessments from live device telemetry. This enables context-aware monitoring of complex CPS environments while preserving transparency and trust. Experimental evaluation using a representative CPS testbed and MITRE ATT\&CK campaign-inspired attack scenarios demonstrates sub-millisecond twin synchronization latency and up to 21.5\% faster threat detection compared with deterministic reasoning alone. The results show that semantic modelling, semantic enrichment, and hybrid reasoning improve, explainability, and resilience without extra system overhead. HySecTwin provides a lightweight, containerized, and extensible framework for secure-by-design digital twin deployments in mission-critical infrastructures.}

\keywords{digital twins, semantic modeling, cybersecurity, machine reasoning}



\maketitle

\section{Introduction}\label{Intro}

Securing cyber–physical systems (CPS) that underpin critical infrastructure is inherently complex. Unlike enterprise IT networks, industrial CPS environments comprise heterogeneous devices and protocols, embed implicit context in physical processes, and exhibit opaque causality between cyber- and physical events. These factors make it difficult to interpret raw signals and pinpoint attack root causes using traditional security tools. A malfunctioning sensor reading, for example, might only be meaningful when correlated with process state and control logic, and subtle attack-induced deviations may be obscured by noisy operational data. This complexity demands security approaches that can integrate diverse data sources and domain knowledge to infer threats in a contextualized, explainable manner.

\textit{Digital twins \textbf{(DTs})}, virtual counterparts of physical assets and processes (physical twins), have emerged in engineering as a means to mirror system state and behavior in real time. Originally used for system design, simulation, and predictive maintenance, DT technology is now increasingly being applied to cybersecurity analysis. By providing a high-fidelity digital mirror of the CPS, a DT can serve as a testbed for detection and response strategies without risking disruption to the live system. Early studies have leveraged DTs for anomaly detection and process fault identification in manufacturing settings (e.g., Gaikwad et al. integrate thermal simulations and sensor data via a DT to detect additive manufacturing faults ~\cite{Gaikwad2020}). In industrial control contexts, DTs have been proposed for monitoring system performance and flagging deviations: for instance, Masi et al. introduce a “cybersecurity digital twin” architecture to secure critical infrastructure~\cite{Masi2023}, and Varghese et al. develop a DT-based intrusion detection system for industrial control systems (ICS)~\cite{Varghese2022}. These works demonstrate that DTs can reduce deployment costs and improve system analysis by enabling detection of incipient failures or attacks before they escalate. Indeed, DTs have been used to support predictive maintenance and anomaly detection in smart factories, and even to emulate attacks in a controlled virtual environment as a means to test countermeasures safely~\cite{Eckhart2018}. However, the use of DTs as dedicated cybersecurity assets remains comparatively niche, and significant gaps must be addressed before their full potential in security operations can be realized \cite{Holmesa}. Prior research surveys the landscape of DT applications in security (e.g., use cases and requirements for cyber-defense~\cite{Eckhart2023}) and identifies opportunities for DTs to augment defense-in-depth strategies~\cite{Holmesa}. A recurring theme is the need to bridge the vocabulary and context between operational technology (OT) engineers and cybersecurity analysts so that a DT can seamlessly support threat monitoring in terms both communities understand.

Crucially, for a DT to act as an effective cybersecurity tool, it must go beyond a basic data-synchronization replica of the physical system. The twin needs to expose semantic context about assets, information flows, and security policies. Without machine-interpretable context, a DT risks becoming a complex but opaque copy of the CPS, one that is no more transparent to security analysis than the original system. In other words, a conventional DT that only mirrors raw states provides an insufficient basis for automated threat analysis, since the burden of interpreting those states (e.g., which sensor readings indicate unsafe operating conditions, which network commands violate policy) falls entirely on human experts or ad hoc scripts. This limitation is evident in existing DT-based security prototypes that lack semantic enrichment: they can replay events and detect certain anomalies, but they struggle to explain why an anomaly is important or to correlate it with higher-level tactics or impacts. As Holmes et al. observed, the integration of DT technology has often insufficiently explored the risks and lacks systematic methods to represent security knowledge~\cite{Holmesa}. Semantic modeling has therefore been identified as a key enabler to unlock the cybersecurity value of DTs~\cite{Eckhart2023}.
% and \cite{Kharlamov2018}. 
By annotating a DT with a formal, machine-interpretable ontology of the CPS (using standards such as the \textit{Resource Description Framework (RDF)} and the \textit{Web Ontology Language (OWL)}), one can encode the relationships and properties that are implicitly understood by engineers, for example, which sensors influence which actuators, what physical process parameters are considered safe, or how network zones and access controls are structured. Such an ontology provides a shared vocabulary and formal conceptualization to integrate heterogeneous data and tools, ensuring that diverse components of a CPS share a common understanding of key concepts like alarms, flows, or authentication. Semantically enriched DTs (sometimes termed 'cognitive twins' or semantic digital twins) have been proposed as a foundation for security monitoring in Industry 4.0 settings. Kharlamov et al.~\cite{Eckhart2018} argue that ontology-backed DTs can greatly simplify analytics and diagnostics by enabling interoperability and automated reasoning over the aggregated knowledge of the system. In the absence of semantic enrichment, any security assessment using a DT would require custom integration for each new data source or rule, and would lack the ability to generalize or reuse knowledge across systems---a clear barrier to scaling security solutions for the Internet of Things (IoT) and industrial CPS. In short, semantic modeling suffuses the DT with understanding of the CPS's structure and context, which is a prerequisite for performing higher-level security analysis in an explainable and automated fashion.

Even with a semantically enriched twin, effective cyber-defense requires reasoning mechanisms that can draw actionable insights from the twin’s knowledge in real time. Recent work on \textit{Security Orchestration, Automation and Response (SOAR)} has started to incorporate DTs as central data hubs~\cite{Empl2022}, but these have so far relied on conventional rule engines or simple analytics. We posit that a hybrid reasoning approach, combining deterministic logic with fuzzy inference, is needed to fully leverage the rich semantic data for cybersecurity~\cite{Sikos2018}. Deterministic reasoning (e.g., a Rete-based rule engine or description-logic reasoner) excels at applying crisp rules and constraints: for instance, enforcing known safety limits, detecting explicit policy violations, or checking invariant conditions in the CPS (such as a valve position that contradicts a sensor reading). It provides explainable conclusions because each alert can be traced to a specific rule or ontological violation, aligning well with expert knowledge and regulatory requirements. Prior DT-based intrusion detection studies have indeed utilized rule-based engines or specification checking to flag unauthorized system states~\cite{Eckhart2018, Varghese2022}. However, purely deterministic logic may fail to catch incipient or ambiguous attacks that do not cleanly violate any single rule, for example, a slow-drifting sensor spoofing or a combination of minor anomalies that collectively indicate malicious intent. Fuzzy reasoning addresses this gap by evaluating degrees of truth and combining multiple soft indicators. By using fuzzy logic, the system can reason about partial rule satisfaction or incorporate heuristic knowledge, thereby recognizing subtle patterns such as gradual performance degradation or intermittent fault symptoms that a crisp rule might overlook. Fuzzy inference has been applied in CPS security to assess noisy signals and produce confidence-weighted alerts (e.g., treating sensor deviations not as binary events but as members of fuzzy sets like “slightly abnormal” vs. “severely abnormal”). In the context of a DT, fuzzy rules can augment the detection of complex attack scenarios – for example, a combination of small pressure fluctuations and moderate network latency might together raise a high suspicion of a stealthy MITM attack, even if each metric alone stays within acceptable bounds. By building a hybrid reasoning architecture, this approach allows us to benefit from the strengths of each: the deterministic component ensures formal compliance and catches obvious violations with zero false positives, while the fuzzy component provides resilience to uncertainty and minor variations, thus reducing false negatives. Notably, our approach does not preclude the incorporation of data-driven machine learning (ML) methods on DT telemetry (as explored by Gaikwad et al.~\cite{Gaikwad2020} and others), but we focus on approaches that retain transparency. Pure ML techniques can identify complex patterns in sensor data, yet they often behave as black boxes and lack the causal explanations necessary for operator trust and forensic analysis~\cite{Eckhart2023}. In contrast, a hybrid reasoning engine grounded in semantic knowledge can produce alerts that are accompanied by human-understandable justifications (e.g., referencing the violated policy or the combined weight of multiple anomalous factors), thereby supporting the explainability required for critical infrastructure defense.

In summary, the problem addressed in this paper is how to secure industrial cyber–physical systems by leveraging DT technology augmented with semantic modeling and hybrid reasoning. We aim to enable a DT-based security framework that can automatically detect and interpret both straightforward and nuanced threats in CPS operations, and do so in a way that is interpretable to humans and aligned with existing cybersecurity frameworks. Our solution approach is grounded in established best practices. For example, it aligns with the NIST Cybersecurity Framework (CSF) core functions \textit{Detect}, \textit{Respond}, and \textit{Recover}~\cite{nistcsf20} by facilitating timely detection of incidents, automated response actions, and system recovery guidance. By tagging detected events with semantic metadata, the approach also makes it possible to map technical findings to high-level CSF categories (e.g., identifying which \textit{Detect} subcategory a particular alert falls under), which can aid in compliance and reporting. This practical orientation ensures that our research outcomes can be readily interpreted in the context of standard security management processes, bridging the gap between theoretical advances and real-world adoption.

\paragraph{Contributions.}
Building on the above insights, this paper presents \textbf{HySecTwin}, a framework for semantic twin-driven security analysis in CPS. The key contributions of our work are as follows:
\begin{itemize}
    \item \textbf{Semantic CPS Digital Twin:} We design an ontology-based semantic model for the cyber-physical system and integrate it with the DT. This semantic annotation of the DT captures security-relevant knowledge about the system’s assets, network topology, physical processes, and policies in a machine-readable format (using RDF/OWL). It provides a unified, extensible representation that enables interoperability and context-sharing across CPS components and security tools.

    \item \textbf{Hybrid Reasoning Engine:} We develop a hybrid reasoning mechanism that combines deterministic rules with fuzzy inference to automatically analyze the semantically enriched twin. The deterministic component (using a rule engine) checks for crisp violations of safety or security conditions (e.g., threshold exceedance, state inconsistencies, known attack signatures), while the fuzzy component evaluates partial evidence and aggregates multiple weak signals to detect complex or stealthy attack patterns. The reasoning engine produces explainable alerts with confidence scores, thus supporting both \emph{explainability} and \emph{early detection} of threats that purely rigid or purely data-driven approaches might miss.

    \item \textbf{Reference Implementation:} We implement a reference prototype of the proposed framework, integrating off-the-shelf and custom components. In our implementation, an open-source DT platform is coupled with a semantic triple store and reasoning engines (a Drools-based expert system and a Fuzzy Logic module) to form a cohesive security twin environment. We outline the engineering decisions and architecture of this prototype, which can serve as a blueprint for deploying semantic security twins in practice. The implementation demonstrates feasibility on contemporary IoT/CPS infrastructure and is extensible to accommodate additional reasoning modules or data sources.

    \item \textbf{CPS Case Study and Evaluation:} We validate the HySecTwin framework on a representative industrial control case study. Using a laboratory CPS testbed, we simulate attack scenarios inspired by the MITRE ATT\&CK for ICS knowledge base (specifically the Dragonfly 2.0 APT campaign). We show that our hybrid reasoner can successfully identify and contextualize the attacks, achieving improved detection coverage compared to a purely deterministic approach. We evaluate system performance under realistic load conditions, measuring the end-to-end detection latency and throughput to ensure the approach meets real-time operational requirements. Furthermore, we demonstrate how each detected event is mapped to the NIST CSF functions and categories, illustrating the framework’s practical relevance for cybersecurity governance.
\end{itemize}

To the best of our knowledge, this work is the first to integrate semantic modeling with a hybrid rule/fuzzy reasoning engine in a DT for ICS security, and to report an evaluation of such a system under CPS workload conditions.

The remainder of this paper is organised as follows. Section~\ref{sec:relatedwork} reviews related work on cybersecurity applications of DTs and highlights the unresolved challenges. Section~\ref{sec:3-Background} formalises the problem and introduces key concepts, including the threat model and semantic modeling foundations. Section~\ref{sec:4-HybridFramework} details the architecture of the proposed HySecTwin framework, while Section~\ref{sec:5-usecase_study-framework_experimentation} describes the implementation and a representative CPS setup used for experimentation. In Section~\ref{sec:6-evaluation}, we present the results of our evaluation, including detection accuracy and performance metrics, followed by a discussion of findings and limitations in Section~\ref{sec:7-discussion}. Finally, Section~\ref{sec:8-Conclusion} concludes the paper and outlines directions for future work.

\section{Related Work}\label{sec:relatedwork}

DTs are a mature technology in engineering and manufacturing and are increasingly studied for cybersecurity. Foundational work defines DTs as automated, bi-directional systems operating at machine speed with continuous synchronization between physical assets and their digital counterparts~\cite{Kritzinger2018c,Eckhart2023}. In cybersecurity, DTs are commonly used as safe proxies for attack emulation and countermeasure testing without disrupting operational systems~\cite{Masi2023,Talkhestani2018}. A closely related research stream originates from predictive maintenance, where anomaly detection identifies early deviations in system behavior preceding faults. Similar mechanisms are now applied to cybersecurity scenarios, where abnormal behavior results from adversarial activity rather than gradual component degradation. Several studies demonstrate anomaly and intrusion detection in cyber–physical systems using DT telemetry or specification-based models. Gaikwad et al.~\cite{Gaikwad2020} show that DT integration in additive manufacturing environments enables anomaly detection through machine-learning analysis of IoT sensor metadata, while Balta et al.~\cite{Balta2019} employ Signal Temporal Logic (STL) specifications within DT models to detect violations in CPS behavior. Such approaches demonstrate that DTs can capture abnormal operational patterns, including attacks that intentionally induce physical faults through process manipulation or control tampering.

Complementary work focuses on predictive modeling of CPS signals. Kummerow et al.~\cite{Kummerow2019} introduce dynamic digital models capable of identifying abnormal behavior by comparing simulated and historical signal trends. Datta et al.~\cite{Datta201714} extend this idea through large-scale “digital duplication,” where swarms of DT instances analyse patterns supporting fault and intrusion detection. While effective for behavioral monitoring, these approaches primarily rely on signal comparison or statistical analysis rather than structured semantic reasoning.

Intrusion detection using DT specifications has also been explored by Eckhart and Ekelhart through CPS Physical Twinning, which treats system inputs and runtime behavior as security-relevant indicators and detects intrusions by comparing observed signals with specification-derived DT behavior~\cite{Eckhartr201861}. Subsequent extensions enable DT generation from CPS specifications and introduce capabilities such as rule-based monitoring, visual alerts, and record-and-replay recovery mechanisms~\cite{Eckhart2018,Eckhart2019}. Although these approaches provide explainable rule-based detection, they largely rely on specification matching and do not integrate semantic knowledge across heterogeneous CPS assets.

A complementary research direction argues that effective security-oriented DTs require \emph{semantic}, machine-readable vocabularies such as RDF and OWL to unify heterogeneous assets, data flows, and policies. Semantic modeling enables interoperability, explainability, and tool reuse by providing a shared representation layer across CPS components~\cite{Kharlamov2019,Eckhart2023}. Without such semantics, DTs struggle to support systematic policy reasoning or automated decision-making. Accordingly, ontologies and knowledge graphs are increasingly adopted to enhance DT interoperability and reasoning. Systematic studies highlight both the promise of semantic technologies and the lack of standardised modeling practices and operational reasoning integration in current DT implementations. Knowledge graph approaches unify heterogeneous DT data sources and enable inference and contextual enrichment, yet are often evaluated in manufacturing or infrastructure scenarios rather than cybersecurity applications.

In security-focused contexts, semantic DT frameworks such as Digital Twin-based Security Analytics (DT2SA) integrate ontological structures, behavioral models, and security analytics to generate shareable cybersecurity knowledge for IoT ecosystems. These frameworks provide valuable architectural contributions but remain primarily architectural or microservice-oriented and do not evaluate reasoning performance under realistic CPS workloads~\cite{Empl2023}. Similarly, federated semantic DT approaches combining DTs, ontologies, and rule-based reasoning point toward integrated security automation pipelines but remain conceptual and lack empirical validation at operational scale~\cite{10431655}.

Beyond direct security applications, the Semantic Sensor Web literature demonstrates how the semantic annotation of sensor and CPS data enables expressive representation and formal analysis of heterogeneous signals, offering potential synergies for DT telemetry enrichment and CPS semantic integration. However, practical integration of such semantic layers with DT reasoning engines for real-time security analysis remains underdeveloped. Reasoning over DT knowledge typically follows three approaches. Deterministic rule engines provide explainable analysis over structured facts~\cite{Eckhartr201861,Eckhart2018,Eckhart2019}, while fuzzy inference supports graded judgments under uncertainty or noisy signals~\cite{Balta2019}. Machine-learning methods analyse DT telemetry to detect behavioural patterns and anomalies, though they often lack causal explainability and operate on data without semantic categorization~\cite{Gaikwad2020,Balta2023}. Consequently, most existing studies employ these techniques independently rather than combining them within unified hybrid reasoning architectures.

\begin{table*}[!t]
\centering
\caption{Comparative analysis of related works against HySecTwin.}
\label{tab:dt_related_comparison}
\footnotesize
\setlength{\tabcolsep}{6pt}
\renewcommand{\arraystretch}{1.08}
% \begin{tabular}{p{2.5cm}ccccc}
\begin{tabular}{p{2.5cm}ccccc}
\toprule
\textbf{Related Works} & \textbf{Semantic Model} & \textbf{Rule Reasoning} & \textbf{Hybrid Reasoning} & \textbf{Explainable} & \textbf{Perf. Eval.} \\
\midrule

Balta et al.~\cite{Balta2019}
& N & N & N & P & P \\

Balta et al.~\cite{Balta2023}
& N & P & N & P & P \\

Eckhart ~\cite{Eckhartr201861,Eckhart2019}
& N & Y & N & Y & N \\

%Eckhart et al.~\cite{Eckhart2019}
%& N & Y & N & Y & N \\

Empl et al.~\cite{Empl2023}
& Y & Y & N & Y & N \\

Varghese et al.~\cite{Varghese2022}
& N & Y & N & P & P \\

\textbf{HySecTwin}
& \textbf{Y} & \textbf{Y} & \textbf{Y} & \textbf{Y} & \textbf{Y} \\

\bottomrule
\end{tabular}

\vspace{1mm}
\begin{flushleft}
\footnotesize
Legends: \textit{Y = Yes, N = No, P = Partially done.}
\end{flushleft}
\end{table*}

\paragraph{Gap and Positioning.}
Despite advances in DT-based monitoring and semantic technologies, existing research lacks a comprehensive DT pipeline that simultaneously (i) semantically encodes the security-relevant CPS structure in a unified, machine-readable form and (ii) integrates hybrid reasoning, combining deterministic (Rete-based) rules with fuzzy inference evaluated under CPS-style workloads. As summarized in Table~\ref{tab:dt_related_comparison}, prior studies address selected capabilities, whereas HySecTwin unifies semantic modeling and hybrid reasoning of CPS twins for security analysis. This work addresses this gap by presenting a semantically grounded DT substrate integrated with hybrid reasoning engines and evaluating their performance under reproducible security-focused scenarios. This enables the system to execute automated reasoning over CPS operational behavior rather than relying solely upon rule-based anomaly detection.

\section{Background}
\label{sec:3-Background}

Building on the problem formulation and research gaps identified in Sections~\ref{Intro} and~\ref{sec:relatedwork}, this section consolidates the conceptual foundations required for the proposed framework. 

\begin{itemize}

    \item \textbf{Physical twin (PT):}
within CPS environments, the PT represents the operational CPS composed of interconnected IoT-enabled devices, including sensors, actuators, controllers, and communication interfaces deployed in critical infrastructures. These IoT components continuously sense environmental and operational conditions (e.g., illumination, temperature, humidity) and exchange telemetry through lightweight communication protocols, enabling real-time monitoring and control. The~\textbf{\textit{PT}}, therefore, captures the actual runtime state of the CPS and constitutes both the primary source of operational telemetry and the principal target of adversarial manipulation, including remote-access abuse, command execution, configuration tampering, and logical–physical inconsistencies. Effective cybersecurity monitoring must, therefore, interpret behavioral deviations jointly across cyber interactions and physical process contexts.

    \item \textbf{Digital twin:}
%(\textbf{\textit{DT}}) 
a digital twin 
is a dynamic virtual representation of the \textbf{\textit{PT}} of a CPS that remains synchronized with its operational counterpart through continuous bidirectional data exchange, enabling monitoring, simulation, state analysis, and decision support without directly affecting physical operations~\cite{10431655, Grieves2016}.

    \item \textbf{Cybersecurity risks:}
complex CPS environments comprise heterogeneous devices, distributed control components, and tightly coupled cyber–physical processes, making them attractive targets for adversarial manipulation. Attackers may exploit remote access paths, alter device configurations, disrupt control logic, or induce logical–physical inconsistencies, complicating root-cause attribution due to operational complexity and noisy system behavior. DTs support cybersecurity monitoring by enabling unified observation and analysis of distributed behavior without affecting operations~\cite{Eckhart,Kharlamov2019}. However, \textbf{\textit{DT}}-enabled monitoring also introduces twin-specific risks: adversaries may manipulate telemetry to induce incorrect twin states, exploit mismatches between physical and digital behavior, or misuse \textbf{\textit{DT}} interfaces to expand attack reach across interconnected components. These challenges motivate Security Digital Twin perspectives that embed explainability and trustworthy interpretation mechanisms as essential for reliable operational decision-making~\cite{Moser2023,SARKER2024935,10431655}.

    \item \textbf{Semantic representation of CPS state:}
within DT-enabled CPS pipelines, operational device data are transformed into structured, machine-interpretable representations. Semantic modeling encodes device attributes, relationships, and runtime conditions as assertions~\cite{Bromander201674,Sikos2019,Sikos2015}, enabling reasoning engines to work over fact bases rather than raw sensor values. These representations promote interoperability across heterogeneous CPS components and ensure consistent interpretation of operational behavior within the DT~\cite{Kharlamov2019}. The resulting semantically enriched system state forms a fact base $\mathcal{F}$ that captures device conditions and relevant process context for cybersecurity interpretation. On this foundation, \textbf{\textit{DT}}-enabled security analytics support intelligent reasoning to monitor behavior and detect anomalies under evolving conditions. Deterministic rules and fuzzy inference systems offer explainable interpretations of deviations, while newer decision-support approaches combine data-driven reasoning to improve situational awareness. Recent work underscores the role of semantically enriched Security Digital Twins in design-time and run-time security analytics~\cite{10431655} and highlights explainable AI for trustworthy, interpretable \textbf{\textit{DT}}-driven cybersecurity decisions~\cite{SARKER2024935}. Integrating deterministic and data-driven reasoning in DT operations enables coordinated anomaly detection, threat diagnosis, and mitigation across complex CPS environments.

    \item \textbf{Digital twin and security reasoning:}
maintaining CPS security in dynamic environments requires reasoning beyond rigid rule-based criteria. DTS enable unified observation of CPS behavior and support anomaly detection by analysing operational states with deterministic rules and hybrid inference. To formally relate DT observations to cybersecurity outcomes, the DT is treated as a structured evidence substrate capturing time-indexed system observations. Let $\mathbf{o}_t$ denote \textbf{\textit{PT}} and \textbf{\textit{DT}} observations at time $t$ (e.g., sensor readings, device states, network reachability, control actions). A semantic lifting function $\phi(\cdot)$ maps these observations to a machine-interpretable fact base $\mathcal{F}_t = \phi(\mathbf{o}_t)$, which is evaluated against a rule base $\mathcal{R}$ via deterministic inference $\mathcal{I}_D(\mathcal{F}_t,\mathcal{R})$, yielding explainable detections of explicit violations or logical inconsistencies, consistent with DT-based rule-driven monitoring~\cite{Eckhart2018,Eckhart,10431655,Sikos2020}.

When uncertainty or gradual deviations occur, $\mathbf{o}_t$ is also encoded as a behavioral feature vector $\vec{x}_t$, enabling graded inference via hybrid reasoning $\mathcal{I}_H(\vec{x}_t)$ to support detection under noisy or ambiguous conditions~\cite{Balta2019,Balta2023}. Recent work shows that semantically enriched Security Digital Twins support both design-time and run-time security analytics~\cite{10431655,Sikos2020}, while explainable AI techniques increase trust and interpretability in DT-driven cybersecurity decisions~\cite{SARKER2024935}. Aligning deterministic and data-driven reasoning within DT operations thus enables coordinated anomaly detection, threat diagnosis, and mitigation in complex CPS environments. Semantic DT behavioral models, combined with CPS telemetry and hybrid reasoning, provide the analytical foundation for DT-based cybersecurity monitoring. Section~\ref{sec:4-HybridFramework} builds on these foundations to present an architecture that integrates these components into a deployable CPS security monitoring framework.

\end{itemize}

\noindent
These foundational concepts inform the design of the \textit{HySecTwin} framework, as described in the following section.

\section{HySecTwin: Hybrid Reasoning Security Framework}
\label{sec:4-HybridFramework}

This section presents the HySecTwin framework as the practical realization of concepts introduced in Section~\ref{sec:3-Background}, describing how the Physical Twin (\textbf{\textit{PT}}), Digital Twin (\textbf{\textit{DT}}), semantic modeling, and reasoning components are integrated into a deployable system. The prototype implementation was deployed using containerized services to ensure reproducibility of the experimental environment. The system components, including the DT platform, hybrid reasoning engine, and supporting middleware, were deployed within a controlled test-bed environment designed to emulate a cyber–physical IoT network. This evaluation focuses on demonstrating the feasibility of integrating DT technologies with hybrid reasoning technology for cybersecurity monitoring. While this experimental build does not intend to provide exhaustive benchmarking, it does establish the functional viability of the architecture within a representative CPS environment.

Three development tracks were evaluated for constructing HySecTwin:
\begin{itemize}
%\item \textit{Option 1}: Physical Twin $\rightarrow$ Digi
\item \textit{Option 1}: Physical Twin $\rightarrow$ Digital Twin $\rightarrow$ Semantic Model $\rightarrow$ Reasoning Engine, where semantic representations are derived from CPS-synchronized DT data.
 Twin $\rightarrow$ Semantic Model $\rightarrow$ Reasoning Engine. Semantic models were derived directly from CPS data.
    
\item \textit{Option 2}: DT framework $\rightarrow$ Semantic Model $\rightarrow$ Physical Twin $\rightarrow$ Reasoning Engine. This approach supported rapid prototyping but introduced inconsistencies between virtual and physical assets requiring later correction.
    
\item \textit{Option 3}: Semantic Model $\rightarrow$ DT $\rightarrow$ Physical Twin $\rightarrow$ Reasoning Engine. This ontology-first strategy proved inefficient due to the absence of operational device data needed to ground semantic and functional modeling.
\end{itemize}

Option 1 was adopted, as constructing the DT directly from real CPS telemetry provides an accurate foundation from which semantic models and reasoning rules are derived. Using machine-readable CPS data improves DT fidelity and enables effective integration with the hybrid reasoning engine.

The \textit{HySecTwin} (\textbf{Hy}brid Inference, \textbf{Sec}urity Monitoring, CPS \textbf{Twin} Integration) framework introduces an integrated security intelligence layer that combines semantic modeling, deterministic reasoning, and fuzzy inference to enable real-time monitoring and anomaly detection through bidirectional PT–DT interaction. Specifically, HySecTwin integrates (i) OWL2RL-compliant RDF semantic models for structured CPS monitoring from digital twin, (ii) deterministic reasoning implemented via the \texttt{durable rules} engine, and (iii) uncertainty-aware fuzzy reasoning using the FuzzyLite library.

\begin{figure*}[t]
\centering
\includegraphics[width=1.0\textwidth]{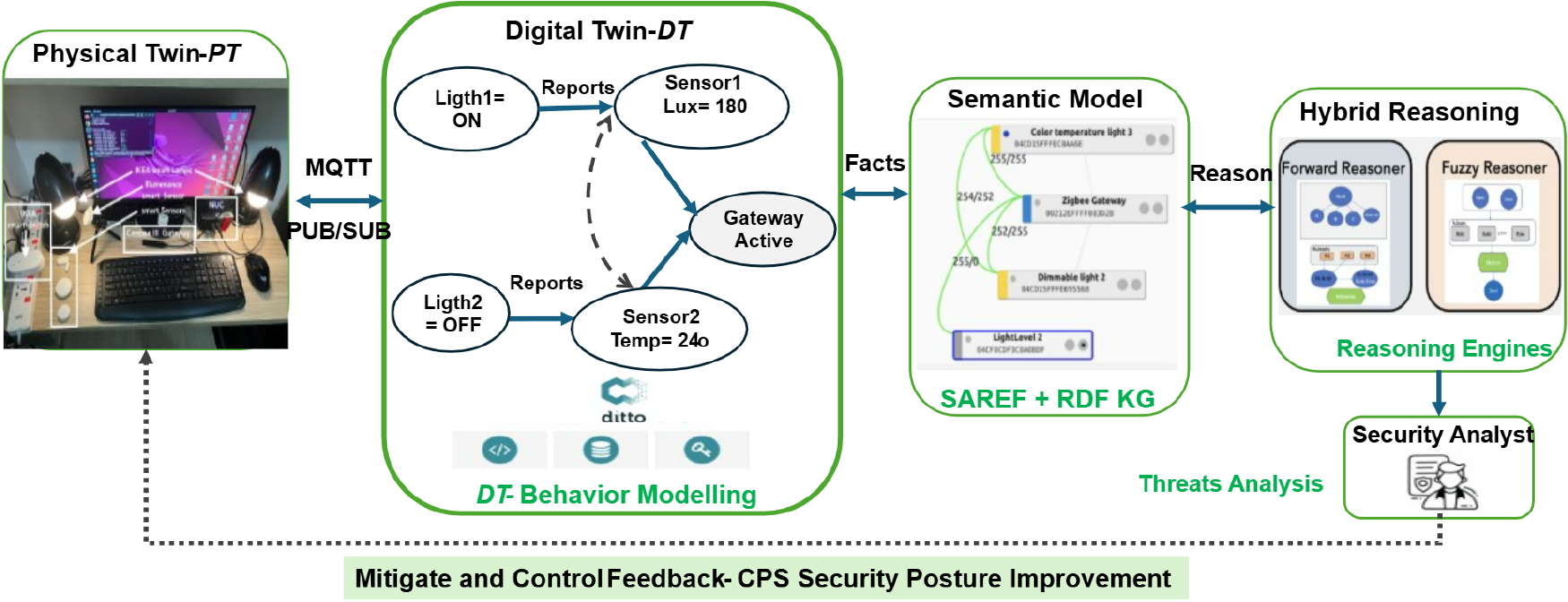}
\caption{HySecTwin framework architecture illustrating physical twin\textbf{\textit{ (PT)}} integration with the digital twin \textbf{\textit{(DT)}}, semantic modeling, and hybrid reasoning engines enabling intelligent cybersecurity decision making for CPS physical twin.}
\label{fig:hysectwin_framework}
\end{figure*}

\subsection{Framework Components}
Following the preferred development sequence illustrated in Fig.~\ref{fig:hysectwin_framework}, we describe the framework components, progressing from physical twin construction to digital twin integration, semantic modeling, and hybrid reasoning.

\subsubsection{CPS Physical Twin}
A CPS \textbf{\textit{PT}} in this work represents a smart lighting CPS composed of multiple interconnected IoT devices forming the real-world operational infrastructure depicted in Fig.~\ref{fig:hysectwin_framework}. These include sensors, actuators, controllers, and communication interfaces that collectively support the digital twin by providing real-time data for control, monitoring, and decision-making, thereby accurately reflecting the system's physical state and behavior. This research simulates a smart building CPS physical twin featuring IoT-based control of lighting, temperature, and humidity IoT devices communicate through a Zigbee mesh connected via IEEE 802.15.4 gateways for real-time data collection~\cite{Gerodimos2023}, in which device telemetry is streamed through an MQTT-based publish-subscribe architecture to enable continuous synchronization with the digital twin.
 
Through this publish-subscribe (PUB/SUB) pipeline, devices continuously stream status updates for real-time monitoring and control. A web-based GUI lets operators adjust thresholds and device states. Sensor nodes start from initial configurations (e.g., lights on) and adapt to local conditions (e.g., activating under low illumination). System behavior arises from both individual device responses and coordinated node interactions. Table~\ref{cps-behaviors} summarises CPS Physical Twin component behaviors under varying operating conditions.

\begin{table}
\centering
\caption{CPS Physical Twin behaviors}
\begin{tabular}{p{0.28\textwidth}p{0.63\textwidth}}\hline
\textbf{CPS Physical Twin}\\  \textbf{(IoT Sensor Nodes and States)} & \textbf{Behavior Description} \\
\hline
Initial states & 
When the CPS PT is powered up or turned on the state of all sensor nodes initialise to the default state.
IoT sensor nodes begin collecting data from its sensors.
The illumination node starts the initial intensity levels \\ \hline
Sensor Nodes & If the IoT sensor node detects low occupancy and low ambient light:
The  sensor nodes turn on to provide sufficient illumination.
If occupancy remains low, but ambient light increases:
The sensor nodes gradually dim to save energy. If the IoT sensor node detects high occupancy:
The  sensor nodes turn on to full brightness \\ \hline
IoT Sensor Nodes  & IoT Sensor nodes (e.g., illumination, temperature, and humidity sensors) continuously collect data (e.g., illumination, temperature, and humidity metadata).
These IoT node sensors then send sensor data to the CPS physical twin for analysis and decision-making. \\
\hline
\end{tabular}
\label{cps-behaviors}
\end{table}

\subsubsection{CPS Digital Twin}
\label{sec:DT}
Our framework uses a CPS DT as a dynamic virtual representation of the PT, which includes IoT devices such as Zigbee lights, sensors and switches within a smart lighting system.

The DT mirrors these devices for real-time monitoring, control, and security analysis. \textbf{\textit{PT}} data is streamed via MQTT 3.1 in a publish–subscribe model, mapping each device to a digital counterpart. The Eclipse Ditto~\cite{EclpliseDITTO} DT platform was chosen for its extensibility, lightweight integration, and strong messaging features, key to our research needs, over Amazon TwinMaker~\cite{AmazonTwinmaker} and Azure DT~\cite{AzureDigitalTwin}. Ditto uses Live and Twin Channels: Live Channel streams real-time data for the reasoning engine, while Twin Channel enables passive synchronization for analysis and threat simulation. Together, these channels create a feedback system integrating operational data with semantic inference, enhancing cyber–physical security.

The DT is generated from a JSON structure that specifies the behavior of each IoT device within the CPS of the corresponding physical twin. These devices have static~\textit{attributes} such as the model, manufacturer, and ID, together with dynamic \textit{features} such as ON/OFF status, brightness, and temperature. These objects emulate physical sensors, enabling semantic inquiries and coordinated behavior. Listing \ref{lst:color-light} presents a node labeled ``Color temperature light 2'' featuring real-time data. 
The DT is composed of: a~\textit{Concierge} for data cohesion; a \textit{Connectivity} module facilitating PT and DT integration through MQTT and gateways; a~\textit{Gateway} ensuring secure REST and WebSocket API communications; and a~\textit{Thing-Search} for efficient indexing and querying. 
These modules support proactive, context-aware control and monitoring in real-time. The JSON format is used to convey changes in device status, enhancing oversight, management, and prompt anomaly detection while supporting pre-rollout testing of updates and security shifts. 

\begin{lstlisting}[
language=json,
firstnumber=1,
basicstyle=\ttfamily\scriptsize,
caption={An extract of ``Color temperature light 2'' node status (CPS Digital Twin)},
label={lst:color-light}
]
"attributes": {
  "project": "Lightswitch"
},
"features": {
  "lights": {
    "2": {
      "colorcapabilities": 16,
      "ctmax": 454,
      "ctmin": 250,
      "etag": "0f1f511fe2958521437c262940734eb0",
      "hascolor": true,
      "lastannounced": "2023-03-26T04:38:24Z",
      "lastseen": "2023-03-26T08:26Z",
      "manufacturername": "IKEA of Sweden",
      "modelid": "TRADFRIbulbE27WSglobeopal1055lm",
      "name": "Color temperature light 3",
      "state": {
        "alert": "none",
        "bri": 36,
        "colormode": "ct",
        "ct": 250,
        "on": true,
        "reachable": true
      },
      "swversion": "1.0.012",
      "type": "Color temperature light",
      "uniqueid": "04:cd:15:ff:fe:c8:aa:6e-01"
    }
  }
}
\end{lstlisting}

\subsubsection{The Semantic Model} 
\label{subsubsec:4.1.3}

Within the HySecTwin framework, the semantic model serves as the formal integration layer linking the physical twin \textbf{\textit{(PT}}), the digital twin (\textbf{\textit{DT}}), and the hybrid reasoning engine. It provides machine-interpretable representations of IoT device attributes and behaviors by transforming CPS telemetry into structured, semantically enriched data suitable for automated reasoning. Device metadata and runtime states are extracted from the CPS network via the ConBee~III gateway and deCONZ, and encoded in JSON format to enable real-time mapping between physical sensor data and corresponding DT entities.
\begin{figure}[hbt]
    \centering
    \fbox{\includegraphics[width=0.45\columnwidth]{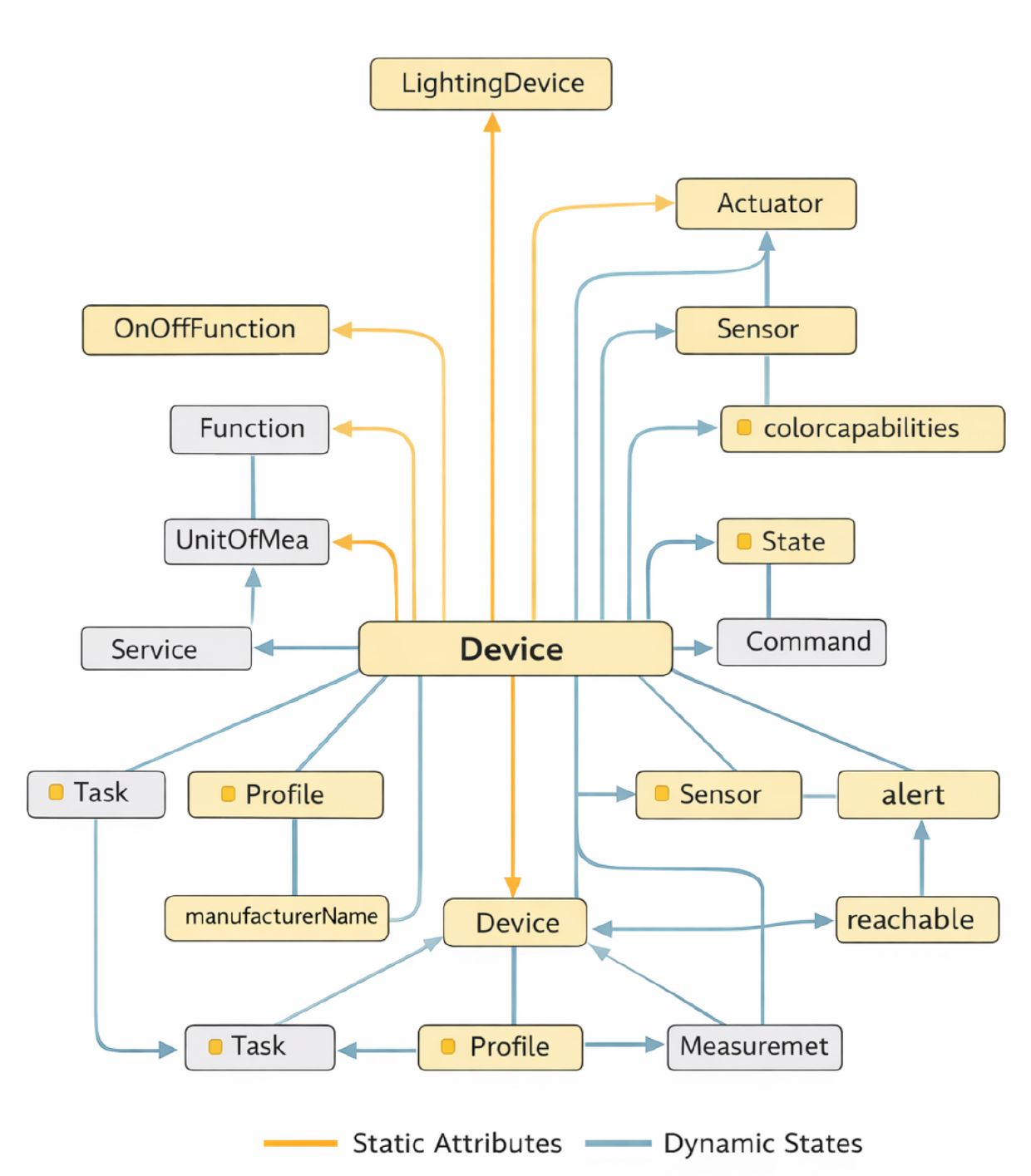}}
    \caption{SAREF-based abstract semantic model representing CPS device relationships enriching Digital Twin.}
    \label{fig2:saref_semantic_model}
\end{figure}
The modeling process follows a data-grounded approach in which operational attributes and state transitions are captured as structured JSON objects. This approach is widely adopted in digital twin ecosystems to support interoperable data integration across domains such as manufacturing, autonomous systems, and cybersecurity~\cite{Alnowaiser2023, Sikos2020, Muralidharanetal2020}. Within HySecTwin, these representations are semantically formalized using the \textit{Smart Applications REFerence Ontology (SAREF)}, which defines a shared vocabulary for CPS entities, including \texttt{Device}, \texttt{Function}, \texttt{State}, \texttt{Property}, and \texttt{UnitOfMeasure}. Where required, lightweight Description Logic extensions are introduced to preserve reasoning tractability within the hybrid inference pipeline~\cite{Horrocksetal2006}.

\begin{figure*}[hb]
\centering
\includegraphics[width=0.9\textwidth]{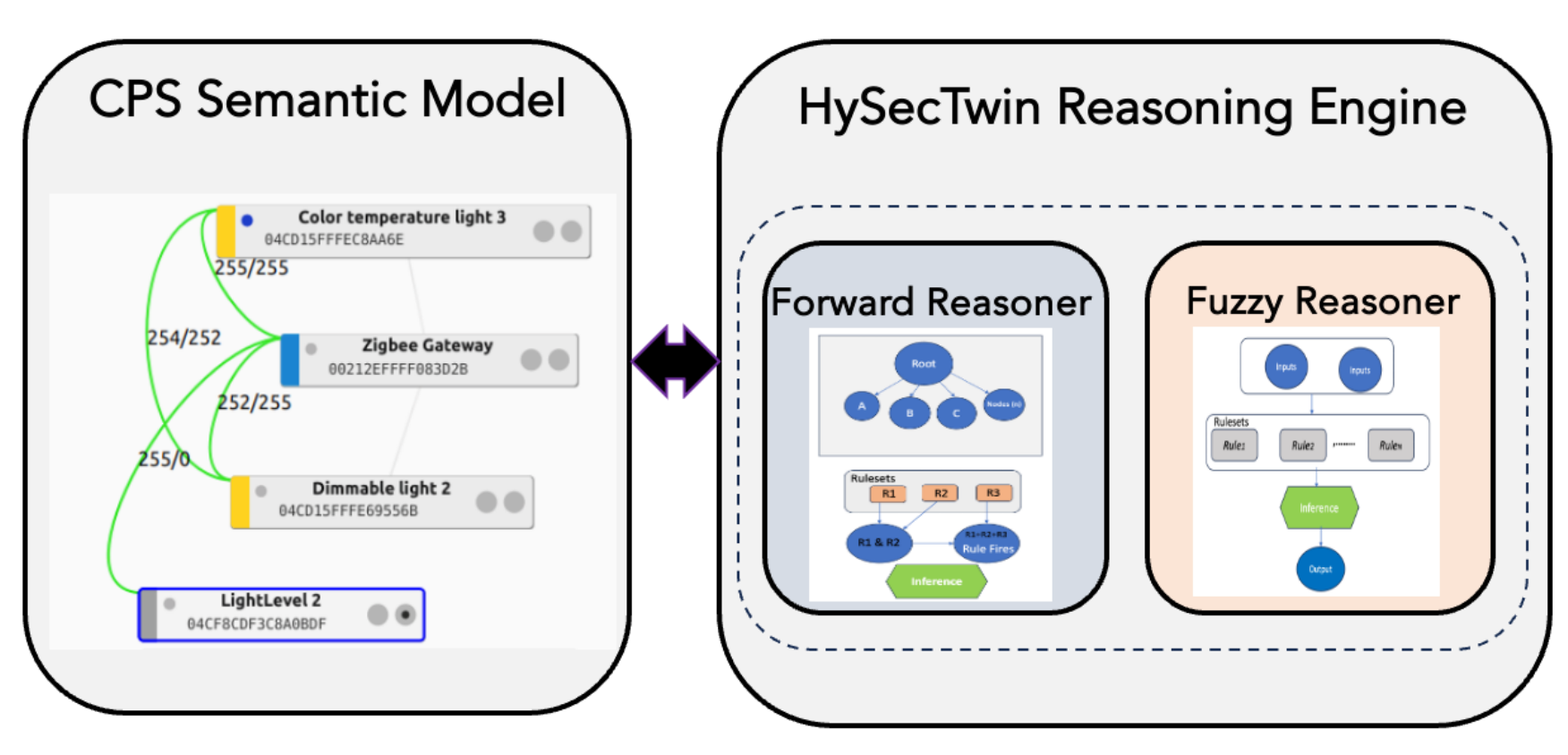} 
\caption{Semantic Model and HySecTwin Reasoner}
\label{fig:3-semantic_model}
\end{figure*}
The semantic layer is implemented as an RDF knowledge graph aligned with SAREF. Static attributes (e.g., \texttt{manufacturerName}, \texttt{modelId}, \texttt{uniqueId}) and dynamic properties (e.g., \texttt{state.on}, \texttt{brightness}, \texttt{temperature}, \texttt{reachable}) jointly constitute the operational fact base underpinning the DT representation in Eclipse Ditto. Figure~\ref{fig2:saref_semantic_model} illustrates a representative ontology segment in which CPS entities such as \texttt{Device}, \texttt{Sensor}, and \texttt{Actuator} are interconnected through functional, state, and measurement relationships derived from SAREF semantics. This structure enables coherent modeling of both static device characteristics and evolving runtime conditions, supporting multi-hop reasoning across device behavior, control actions, and environmental observations. The semantic model maintains synchronized primary (active) and secondary (passive) \textbf{\textit{DT}} views, aligned with the \textbf{\textit{PT}} and mapped to live and twin channels for real-time and historical analysis. By deriving semantics directly from CPS telemetry, the model remains tightly coupled to the physical system while ensuring consistency and interoperability across heterogeneous components. This unified representation provides a coherent knowledge base that enables the hybrid reasoning engine to process streaming data, enforce deterministic rules, and perform fuzzy inference for cybersecurity analytics.

\subsubsection{The Reasoning Engine}
\label{subsubsec:4.1.4-reasoning_engine}

\begin{figure}[!]
\centering
\includegraphics[width=1.0\columnwidth]{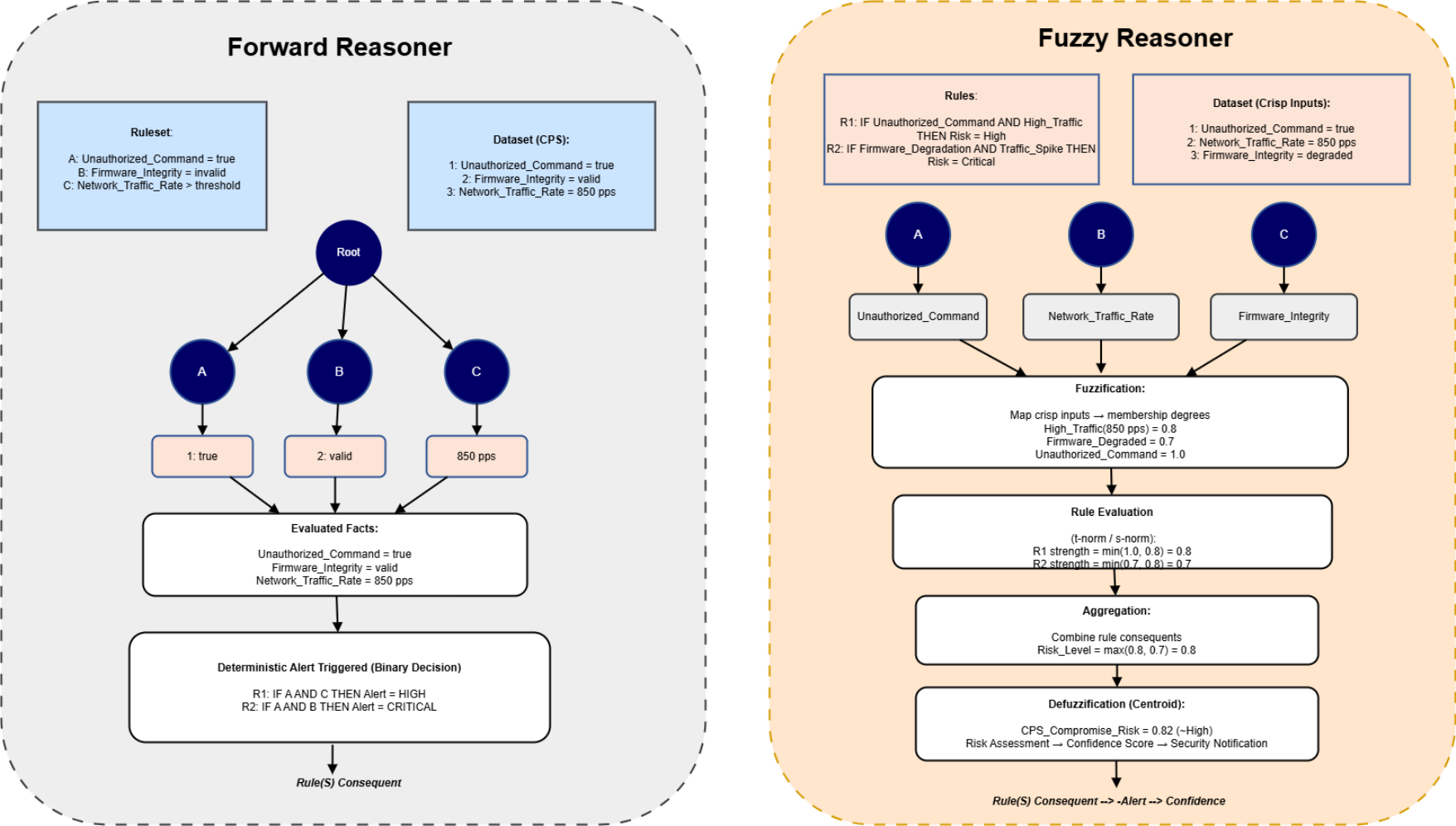}
\caption{Cybersecurity reasoning in HySecTwin: deterministic rule evaluation (left) and hybrid fuzzy-enhanced risk assessment (right).}
\label{fig:fig-5}
\end{figure}

The reasoning engine constitutes the analytical core of HySecTwin, transforming the semantically structured CPS \textit{\textbf{twin}} \textbf{\textit{(PT)}} representation (Section~\ref{subsubsec:4.1.3}) into actionable cybersecurity intelligence. The reasoning engine employs Forward and Fuzzy reasoners, refer to Fig. \ref{fig:3-semantic_model}, integrated with the semantic model. Instead of operating on raw telemetry, the engine consumes a normalized fact base \( \mathcal{F} \), derived from SAREF-aligned semantic assertions exported by the \textbf{\textit{DT}}, consistent with the semantic lifting of CPS observations into machine-interpretable facts. Device states, operational context, and security-relevant attributes are encoded as typed assertions (e.g., \texttt{Unauthorized\_Command = true}, \texttt{Firmware\_Integrity = degraded}, \texttt{Network\_Traffic\_Rate = 850~pps}), enabling consistent reasoning across heterogeneous CPS components. Fig.~\ref{fig:fig-5} presents the two complementary inference mechanisms implemented in HySecTwin: deterministic rule-based reasoning and hybrid fuzzy-enhanced reasoning.

\paragraph{Deterministic Reasoning.}

Deterministic inference is designed to identify explicit security violations through rule-based evaluation of the fact base. HySecTwin employs the \texttt{durable\_rules} framework, implementing a Rete-based forward-chaining mechanism for efficient evaluation over streaming CPS data~\cite{liu2016rule,anicic2010rule}. Rules are defined as follows:

\[
r_i : \text{IF } (C_1 \land \cdots \land C_k) \text{ THEN } A
\]

\noindent where \( C_i \in \mathcal{F} \), aligning with rule-driven \textbf{\textit{DT}} monitoring of CPS behavior, and the inference process is formalized as \( \mathcal{I}_D : \mathcal{F} \times \mathcal{R} \rightarrow \{0,1\} \), producing binary outcomes that indicate policy violations. This enables low-latency and explainable detection of unauthorized state transitions, firmware integrity breaches, and abnormal traffic thresholds.

\paragraph{Hybrid Fuzzy-Enhanced Reasoning.}

To capture uncertainty and gradual deviations inherent in CPS environments, HySecTwin extends deterministic inference with fuzzy reasoning~\cite{rashid2022hybridids, klir1995fuzzy, wang2021faultdiag}. The semantic attributes in \( \mathcal{F} \) are mapped to a feature vector \( \vec{x} \in \mathbb{R}^k \), enabling a graded behavioral representation of CPS conditions. Membership functions transform these values into degrees of truth, $\mu(\vec{x}) \in [0,1]$
, which are aggregated to compute an overall compromise score consistent with hybrid reasoning for uncertain CPS behavior. The resulting inference is defined as follows:

\[
\mathcal{I}_H(\vec{x}) =
\begin{cases}
1, & \mu(\vec{x}) \geq \theta, \\
0, & \text{otherwise}.
\end{cases}
\]

This enables detection of ambiguous or evolving threats, such as progressive firmware degradation or coordinated traffic anomalies, using confidence-weighted assessments. Sharing a common semantic foundation, the two inference modes form a unified reasoning pipeline: deterministic reasoning provides precise, auditable detection of explicit violations, while fuzzy inference increases sensitivity to uncertain, context-dependent behaviors. Together, they create a closed-loop cyber–physical analysis process in which semantically enriched digital twin observations are continuously evaluated to produce interpretable alerts and risk-aware insights, delivering a scalable, explainable, and robust cybersecurity capability aligned with CPS (\textbf{\textit{PT}}) operational constraints.

\section{Case Study: Framework Experimentation}
\label{sec:5-usecase_study-framework_experimentation}

Building on the proposed frmaework defined in Section~\ref{sec:4-HybridFramework}, this section evaluates how the integrated Physical Twin, Semantic Model, \textbf{\textit{Digital Twin (DT),}} and Reasoning Engine operate as an intelligent cybersecurity system. While prior sections describe data capture and representation, here we assess how reasoning transforms CPS telemetry into actionable security insights. HySecTwin is evaluated using campaign-inspired scenarios based on MITRE ATT\&CK C0012 (Dragonfly 2.0)\footnote{\url{https://attack.mitre.org/groups/G0035/}}, a representative CPS-focused threat campaign. The experiments demonstrate detection performance under deterministic and hybrid reasoning modes within a fully containerised testbed.

\subsection{Methods and Measurement Protocol}

\paragraph{Objectives.}
The evaluation is designed to measure: (i) end-to-end latency from message publication to persistence, (ii) internal pipeline latencies, (iii) throughput under controlled load, and (iv) the effect of fuzzy scoring on alert confidence and computational cost.
\paragraph{Experimental Setup.}
The evaluation is conducted in a containerized environment on Ubuntu~22.04, integrating Eclipse Ditto, (MQTT broker), the Durable Rules worker, an optional FuzzyLite scorer, MongoDB, InfluxDB~1.8, and Grafana. MQTT communication \cite{oasis-mqtt-v3_1_1, oasis-mqtt-v5_0}. Synthetic workloads are generated using scripted MQTT publishers and k6 (xk6-mqtt), enabling controlled and repeatable traffic patterns.

\subsection{Performance Evaluation Strategy}

To assess the practical viability of HySecTwin, we evaluate both its baseline efficiency and its behavior under increasing operational stress. The objective is to determine not only how the system performs under nominal conditions but also how it responds when exposed to sustained and intensified telemetry streams.
\newline 
\emph{Benchmarking} establishes reference performance under controlled and repeatable workloads. It characterizes normal system behavior, including latency and throughput under steady-state conditions, and provides a stable baseline against which subsequent experiments can be compared.
\newline 
\emph{Load testing} progressively increases traffic intensity to examine scalability and robustness. This evaluation mode is particularly important in cyber-physical systems, where reliability under concurrent data streams is critical for maintaining operational stability and security monitoring continuity~\cite{nist-sp-800-82r2}. By observing performance degradation trends and saturation points, load testing validates whether the Digital Twin and reasoning pipeline remain dependable under realistic operational pressure.
\newline
\emph{Workloads (including smoke load).} Telemetry replay follows Eclipse Ditto topic conventions: 
\texttt{cps/\textless thingId\textgreater} (active twin channel) and 
\texttt{twin/\textless thingId\textgreater} (persisted twin channel). 
Each message carries a source timestamp (\texttt{ts\_ms}) to enable precise latency measurement across the \textbf{\textit{CPS–DT}}–reasoning pipeline.

Before formal benchmarking, a short \emph{smoke load} is executed at low intensity. This preliminary step verifies end-to-end connectivity, database persistence, and instrumentation accuracy, thereby preventing experimental bias caused by configuration or deployment errors. Once system correctness is confirmed, controlled workload sweeps are performed to measure steady-state latency and throughput under both baseline and stress conditions.

\begin{table}[h!]
\centering
\caption{Technical Workload Parameters}
\label{tab:workload_parameters}
\begin{tabular}{p{3cm} p{4cm} p{4cm}}
\toprule
\textbf{Workload Type} & \textbf{Traffic Characteristics} & \textbf{Technical Parameters} \\
\midrule
Smoke Load & Very low intensity validation & 1--5 VUs, 60--180\,s duration, 1--5\% peak publish rate, Mongo insert success $\geq$99.9\% \\
\addlinespace
Benchmark Baseline & Fixed, controlled traffic & Constant VUs, steady-state execution window, stable publish rate \\
\addlinespace
Load Sweep & Increasing stress profile & Progressive VU scaling, extended duration, increased publish rate \\
\bottomrule
\end{tabular}
\end{table}

\paragraph{Metrics.}

System performance is characterized through stage-wise and end-to-end latency measurements across the CPS–Digital Twin–reasoning pipeline. These measurements allow us to isolate where delays occur and to quantify how long telemetry takes to propagate through the full monitoring loop.

Let {\Large \(t_{\text{src}}\), \(t_{\text{rules}}\)}, and {\Large \(t_{\text{db}}\)} denote the source timestamp embedded in the payload, the time of rule ingestion, and the time of successful database persistence, respectively. We compute three latency components as follows:

{\Large
\[
\Delta_{\text{emqx}\rightarrow\text{rules}}, \quad
\Delta_{\text{rules}\rightarrow\text{db}}, \quad
\Delta_{\text{end}}.
\]
}

These correspond to message ingestion delay, processing-to-persistence delay, and total end-to-end latency. For each configuration, we report p50 and p95 latency (capturing typical and near-worst-case behavior), mean latency, throughput (messages/s), MQTT round-trip time, and MongoDB CRUD performance in a YCSB-style format~\cite{cooper2010ycsb}. Collectively, these indicators reveal both efficiency and operational stability of the reasoning pipeline under benchmark and stress conditions.

\paragraph{Hybrid scoring.}

In Hybrid mode, FuzzyLite returns a confidence value in \([0,1]\), representing the inferred degree of compromise or anomaly. Scores are grouped into \{\textit{low}, \textit{med}, \textit{high}\} bands to support interpretable analysis. This enables evaluation not only of processing speed but also of detection certainty, providing insight into the trade-off between responsiveness, robustness, and explainability in cybersecurity monitoring.

\paragraph{Standards alignment and reproducibility.}

The evaluation methodology aligns with NIST SP~800-82 guidance for industrial control systems~\cite{nist-sp-800-82r2} and the measurement principles of SP~800-55~\cite{nist-sp-800-55r2}. Each configuration is executed three times under steady-state conditions, with consistent parameters, to ensure statistical reliability and reduce variability. This approach strengthens the validity and reproducibility of the reported results.

\subsection{Use Case 1: Deterministic Baseline (C0012)}

The first use case establishes a cybersecurity performance baseline by deploying the \texttt{durable\_rules} engine in isolation under the MITRE ATT\&CK Dragonfly~2.0 campaign (C0012). Dragonfly represents a real-world advanced persistent threat targeting critical infrastructure, including credential harvesting, remote access, and command execution techniques. As such, it provides a realistic benchmark for evaluating rule-based CPS threat detection. Telemetry tagged with \texttt{campaign=C0012} was processed over a 48-hour observation window to simulate sustained operational exposure. The deterministic engine applied crisp, pre-defined rules mapped to campaign behaviors, producing binary detections aligned with known attack signatures. Evaluation focuses on latency statistics (mean, p50, p95, p99), alert volume, and risk-band distribution. From a cybersecurity perspective, this use case measures the system’s ability to rapidly and reliably detect well-defined attack patterns while maintaining low processing overhead. The results establish a reference for response time, detection consistency, and computational efficiency in a rule-governed CPS security context.

\subsection{Use Case 2: Hybrid Engine Across Campaigns}

The second use case evaluates a hybrid reasoning setup combining deterministic \texttt{durable\_rules} with FuzzyLite-based fuzzy inference. Unlike the single-campaign baseline, it introduces diverse adversarial behaviors across four MITRE ATT\&CK campaigns: C0012 (credential-based intrusion), C0020 (multi-stage spearphishing), C0025 (cloud-based data exfiltration), and C0028 (custom malware and C2). Telemetry from these campaigns was ingested concurrently over 72 hours to simulate complex, overlapping cyber threats in CPS (\textbf{\textit{PT}}) environments, including stealthy, staged, and partially observable attacks. The evaluation compares latency, confidence distributions, detection stability, and inference consistency across campaigns and engines. Deterministic rules enable fast, explainable detection of explicit policy violations, while fuzzy inference provides graded confidence for ambiguous, gradual, or multi-signal anomalies. This use case therefore examines detection speed, adaptive threat interpretation, and resilience under heterogeneous attack patterns. The results show that hybrid reasoning improves situational awareness, mitigates blind spots in rigid rule systems, and increases robustness in CPS DT deployments.

\section{Evaluation}
\label{sec:6-evaluation}

This section evaluates the HySecTwin framework using the cyber-physical use cases defined in Section~\ref{sec:5-usecase_study-framework_experimentation}. The evaluation focuses on three dimensions: (i)  CPS-DT performance and latency behavior, (ii) the consistency of digital twin synchronization, and (iii) the effectiveness of the reasoning engine in detecting cyber-physical anomalies mapped to MITRE ATT\&CK campaigns. Quantitative results are reported using latency, throughput, and detection-time metrics. Tables provide the primary record for reproducibility and statistical validity, while figures illustrate distributional trends and comparative behavior across reasoning modes.

\subsection{Benchmark Scope and Use-Case Mapping}

\begin{figure}[!t]
  \centering
  \includegraphics[width=0.65\linewidth]{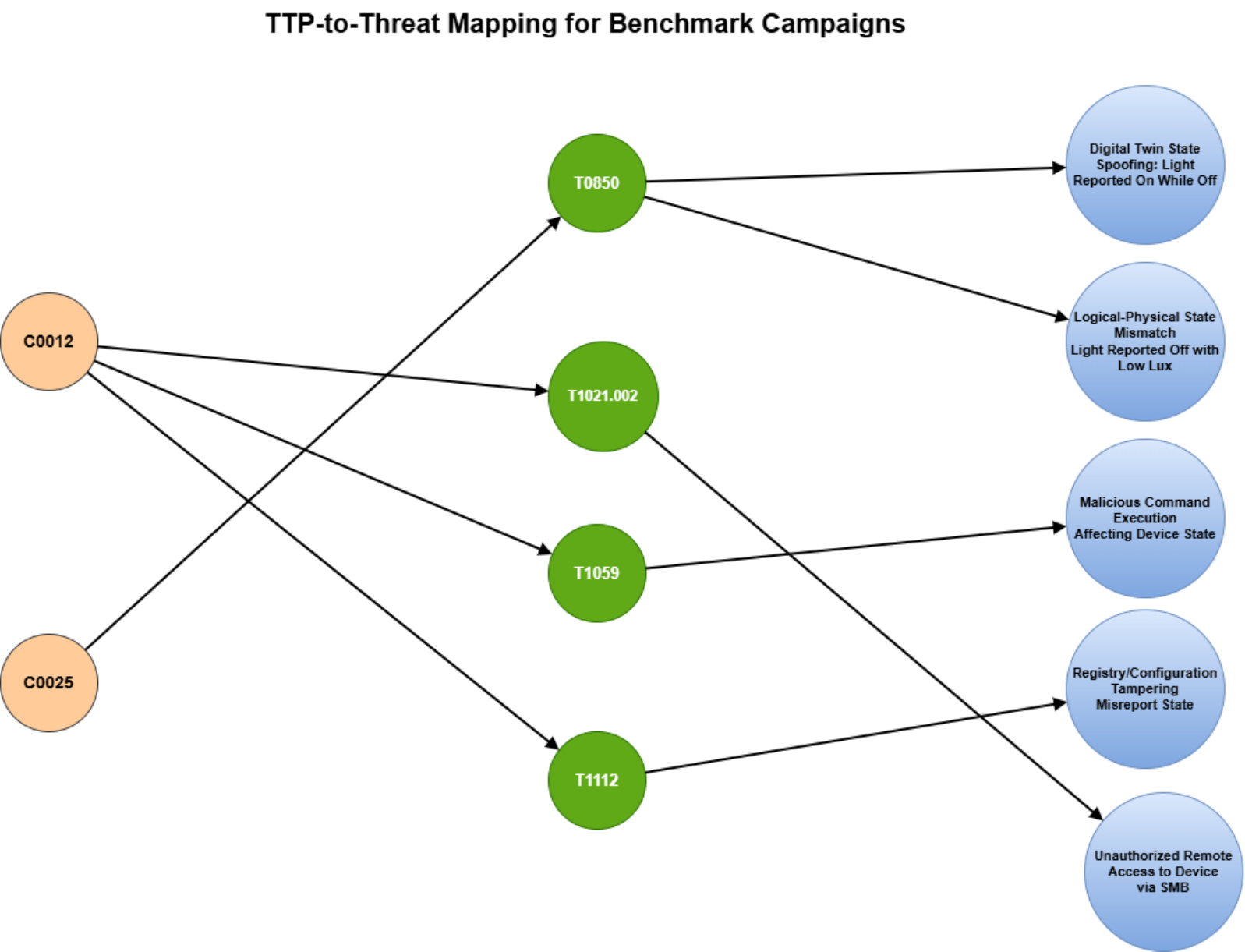}
  \caption{\textbf{TTP to Threat Mapping}}
  \label{fig:12-ttp_to_threat_graph}
\end{figure}

The experimental use cases are grounded in benchmark adversarial campaigns and ATT\&CK techniques mapped to observable threats in the lightswitch CPS domain. These scenarios provide a consistent basis for evaluating nominal and malicious behavior introduced in Section~\ref{sec:5-usecase_study-framework_experimentation}. As shown in Fig.~\ref{fig:12-ttp_to_threat_graph}, campaign \textbf{C0012} (Dragonfly 2.0) includes techniques such as \textbf{T1059}, \textbf{T1112}, and \textbf{T1021.002}, corresponding to malicious command execution, configuration tampering, and unauthorized remote access. Campaign \textbf{C0025} is linked with \textbf{T0850}, representing digital twin state spoofing and logical–physical inconsistencies. Accordingly, \textbf{UC1} captures direct control-oriented attacks with explicit state mismatch effects, whereas \textbf{UC2} captures more complex behaviors including spoofed telemetry, command-and-control-like toggles, logical–physical inconsistencies, and suspicious data transmission. Table~\ref{tab:uc-lightswitch-mapping} and Fig.~\ref{fig:12-ttp_to_threat_graph} summarize these mappings, linking system metrics (latency, throughput) with cybersecurity outcomes (detection time, accuracy, and confidence).

\begin{table}[t]
\centering
\caption{Use-case to lightswitch-domain mapping used to operationalise campaign-inspired behaviors. \emph{Note:} C00xx are ATT\&CK \emph{Campaign} IDs; techniques are referenced separately in the evaluation.}
\label{tab:uc-lightswitch-mapping}

\footnotesize
\begin{tabularx}{\linewidth}{@{}l l X@{}}
\toprule
\textbf{Use case} & \textbf{ATT\&CK object} & \textbf{Lightswitch-domain symptom (observable)} \\
\midrule

\textbf{UC1} & \textbf{C0012} (Dragonfly 2.0)
& \RaggedRight Unauthorised remote control of a light; unexpected command execution; configuration tampering causing \emph{state mismatch} (reported state versus physical context). \\

\textbf{UC2} & \textbf{C0020, C0025, C0028}
& \RaggedRight Mixed behaviors across campaigns: periodic command/control-like toggles (C2 behavior), \emph{logical--physical mismatch} (e.g., ``reported OFF with low lux'' or spoofed twin state), and suspicious data transmission patterns indicative of exfiltration-like behavior (e.g., excessive telemetry bursts or feature writes). \\

\bottomrule
\end{tabularx}
\end{table}

\begin{table}[htbp]
\centering
\caption{Live channel latency (ms).}
\label{tab:4-uc_latency_live}
\begin{tabular}{p{5.65cm}cccccccc}
\hline
\textbf{thingId} & \textbf{N} & \textbf{Mean} & \textbf{Median} & \textbf{P95} & \textbf{P99} & \textbf{Min} & \textbf{Max} & \textbf{Jitter} \\
\hline
live/00-15-8d-00-05-48-e4-7c-01-0006/state & 1  & 0.00 & 0.00 & 0.00 & 0.00 & 0.00 & 0.00 & -- \\
live/00-21-2e-ff-ff-0e-19-2c-01/state     & 1  & 0.00 & 0.00 & 0.00 & 0.00 & 0.00 & 0.00 & -- \\
live/04-cd-15-ff-fe-c8-aa-6e-01/state     & 17 & 0.24 & 0.00 & 1.00 & 1.00 & 0.00 & 1.00 & 0.44 \\
live/04-cd-15-ff-fe-c8-c0-12-01/state     & 17 & 0.41 & 0.00 & 1.00 & 1.00 & 0.00 & 1.00 & 0.51 \\
live/b4-e3-f9-ff-fe-a0-c1-b7-01/state     & 22 & 0.27 & 0.00 & 1.00 & 1.00 & 0.00 & 1.00 & 0.46 \\
live/b4-e3-f9-ff-fe-a6-65-90-01/state     & 30 & 0.40 & 0.00 & 1.00 & 1.00 & 0.00 & 1.00 & 0.50 \\
\hline
\end{tabular}
\end{table}

\begin{table}[t]
\centering
\caption{Twin channel latency (ms).}
\label{tab:5-latency-twin}

\footnotesize
\begin{tabularx}{\linewidth}{@{}X c c c c c c c c@{}}
\toprule
\textbf{thingId} & \textbf{N} & \textbf{Mean} & \textbf{Median} & \textbf{P95} & \textbf{P99} & \textbf{Min} & \textbf{Max} & \textbf{Jitter} \\
\midrule
\RaggedRight twin/...19-2c-01/state & 1  & 0.00 & 0.00 & 0.00 & 0.00 & 0.00 & 0.00 & -- \\
\RaggedRight twin/...aa-6e-01/state & 17 & 0.24 & 0.00 & 1.00 & 1.00 & 0.00 & 1.00 & 0.44 \\
\RaggedRight twin/...c0-12-01/state & 17 & 0.35 & 0.00 & 1.00 & 1.00 & 0.00 & 1.00 & 0.49 \\
\RaggedRight twin/...c1-b7-01/state & 22 & 0.27 & 0.00 & 1.00 & 1.00 & 0.00 & 1.00 & 0.46 \\
\RaggedRight twin/...65-90-01/state & 30 & 0.27 & 0.00 & 1.00 & 1.00 & 0.00 & 1.00 & 0.45 \\
\bottomrule
\end{tabularx}
\end{table}

\subsection{Performance and Load Evaluation}

\begin{figure*}[!t]
\centering
\includegraphics[width=1.0\textwidth]{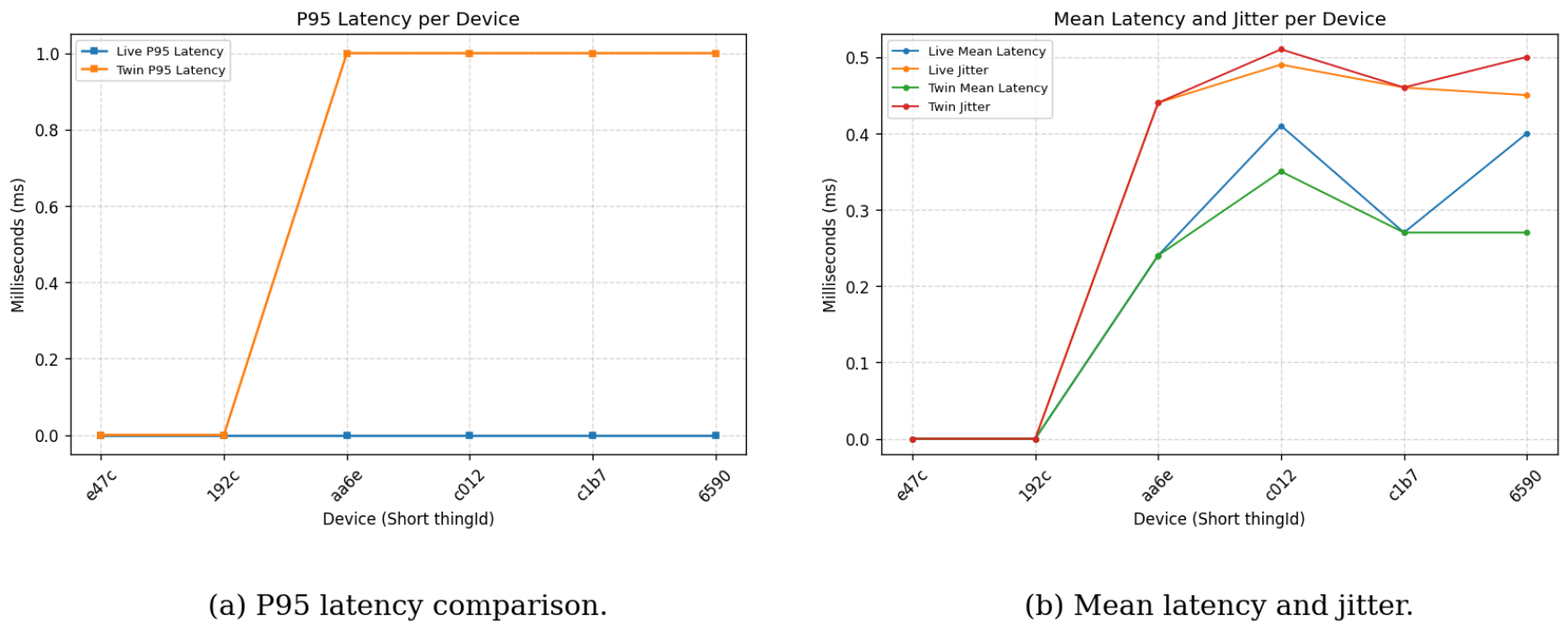}
\caption{Latency and jitter comparison across Live and Twin channels.}
\label{fig:latency-perfm}
\end{figure*}

The framework was evaluated to determine whether continuous CPS (\textbf{\textit{PT}}) monitoring and twin-state updates can be achieved without degrading operational timing behavior. Tables~\ref{tab:4-uc_latency_live} and~\ref{tab:5-latency-twin} report end-to-end latency for the Live (physical CPS) and Twin (digital twin) channels. Across all devices, median latency remains approximately $0$\,ms, while P95/P99 values stay within $1$\,ms with low jitter, indicating deterministic and stable execution under both nominal and adversarial conditions. Figure~\ref{fig:latency-perfm} shows that latency distributions are closely aligned across both channels, confirming that the twin channel preserves the temporal characteristics of the physical process with negligible overhead.

\begin{table}[htbp]
\centering
\caption{Latency summary (Mongo/Influx).}
\label{tab:6-uc_load_testing-mongo-summary}
\footnotesize

\begin{tabular}{l l l r r r r r r r}
\toprule
\textbf{Metric} & \textbf{UC} & \textbf{Channel} & \textbf{Count} & \textbf{Mean} & \textbf{Median} & \textbf{P95} & \textbf{P99} & \textbf{Min} & \textbf{Max} \\
\midrule
\texttt{lat\_end\_ms}       & UC1+UC2 & live & 56940 & 5.00 & 5.00 & 5.90 & 5.98 & 4.0 & 6.0 \\
\texttt{lat\_rules\_db\_ms} & UC1+UC2 & live & 56940 & 3.25 & 3.25 & 3.93 & 3.99 & 2.5 & 4.0 \\
\texttt{mongo\_insert\_ms}  & UC1+UC2 & live & 56940 & 2.00 & 2.00 & 2.80 & 2.98 & 1.5 & 3.0 \\
\texttt{lat\_end\_ms}       & UC1+UC2 & twin & 56940 & 5.10 & 5.10 & 5.98 & 6.05 & 4.0 & 6.0 \\
\texttt{lat\_rules\_db\_ms} & UC1+UC2 & twin & 56940 & 3.30 & 3.30 & 3.95 & 4.00 & 2.5 & 4.0 \\
\texttt{mongo\_insert\_ms}  & UC1+UC2 & twin & 56940 & 2.05 & 2.00 & 2.85 & 2.99 & 1.5 & 3.0 \\
\bottomrule
\end{tabular}

\end{table}

The results collectively show that the Digital Twin synchronization, semantic processing, and reasoning pipeline preserve low-latency operation across both channels. This satisfies industrial control timing expectations, such as those outlined in NIST SP~800--82, where bounded latency and stable temporal behavior are essential for safe and secure operation. For HySecTwin, these findings establish a reliable performance foundation for subsequent analysis of detection accuracy and cybersecurity effectiveness.

\begin{figure*}[!t]
\centering
\includegraphics[width=1.0\textwidth]{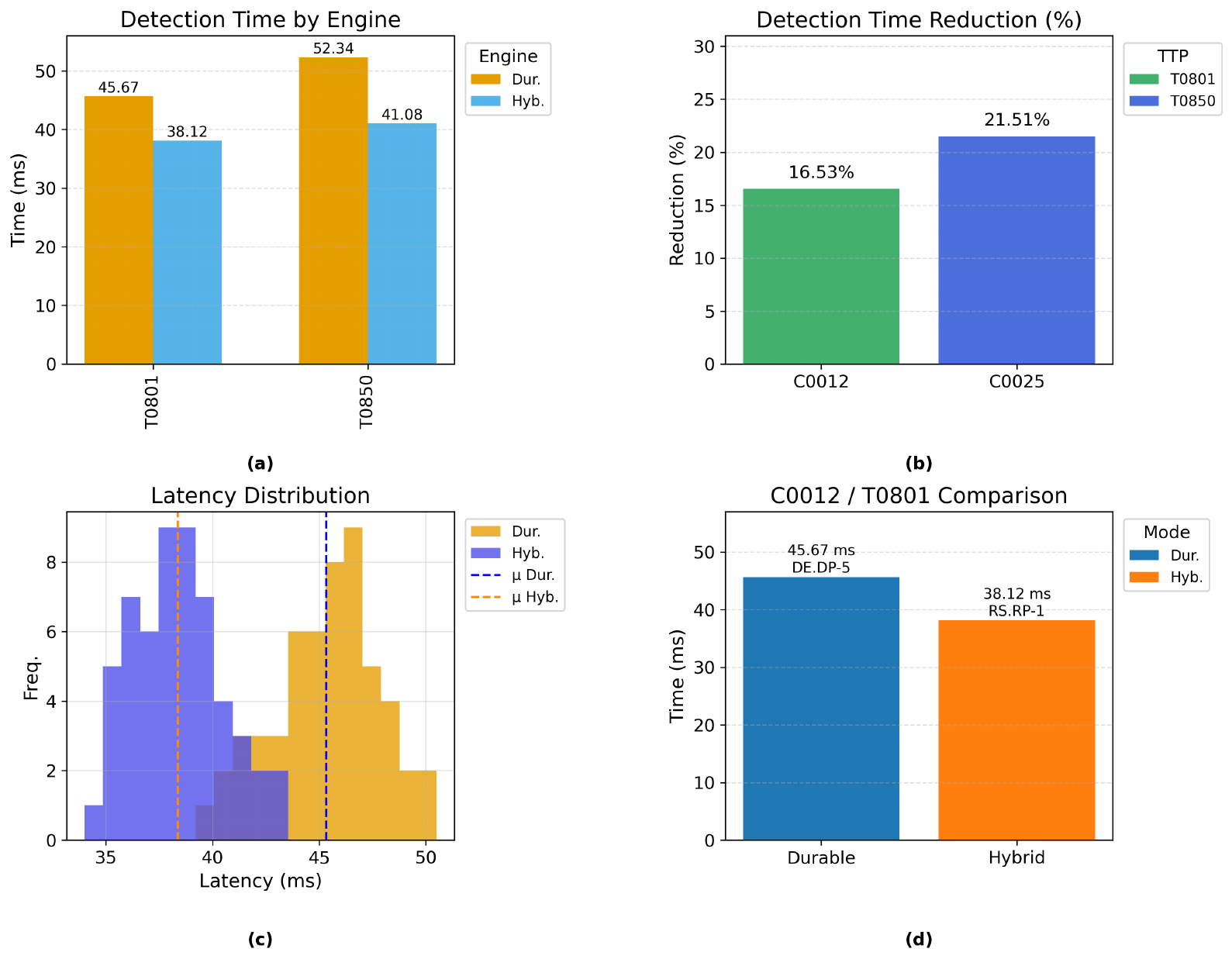}
\caption{HySec-DT threats detection analysis.}
\label{fig:hysecdt_detection_analysis}
\end{figure*}
\subsection{Cyber Threats Detection Analysis}

This subsection evaluates the effectiveness of the reasoning engine in detecting benchmark cyber threats represented through campaign-driven CPS attack scenarios. The analysis compares deterministic (\texttt{durable\_rules}) and hybrid reasoning configurations using detection latency as the primary performance indicator across UC1 and UC2 conditions.
Table~\ref{tab:9-ttp-threat-mapping} presents the mapping between ATT\&CK campaigns, TTPs, and CPS-observable behaviors used as the evaluation ground truth. These mappings operationalize cyber-physical attack semantics and provide a consistent basis for interpreting detection outcomes. Comparative results in Table~\ref{tab:10-ttp-threat-mapping}, annotated with NIST CSF categories, show that the hybrid configuration consistently achieves lower detection latency than the deterministic baseline while preserving stable execution behavior. Table~\ref{tab:11_nist_ttp_detailed} further provides a focused comparison for campaign C0012 (TTP: T0801), where the reduction in detection time is clearly evident.
Figure~\ref{fig:hysecdt_detection_analysis}(a) compares detection times across campaigns and reasoning modes, while Fig.~\ref{fig:hysecdt_detection_analysis}(b) quantifies the percentage reduction achieved by the hybrid engine. Figure~\ref{fig:hysecdt_detection_analysis}(d) provides the detailed comparison for campaign C0012. Latency distributions in Fig.~\ref{fig:hysecdt_detection_analysis}(c) show that the hybrid reasoning engine achieves lower mean detection latency with reduced variance, indicating improved responsiveness in CPS threat detection.

The relative improvement is quantified as:
\[
\text{Improvement (\%)} =
\frac{T_{Durable}-T_{Hybrid}}{T_{Durable}} \times 100
\]

\begin{figure*}[!t]
\centering
\includegraphics[width=1.0\textwidth]{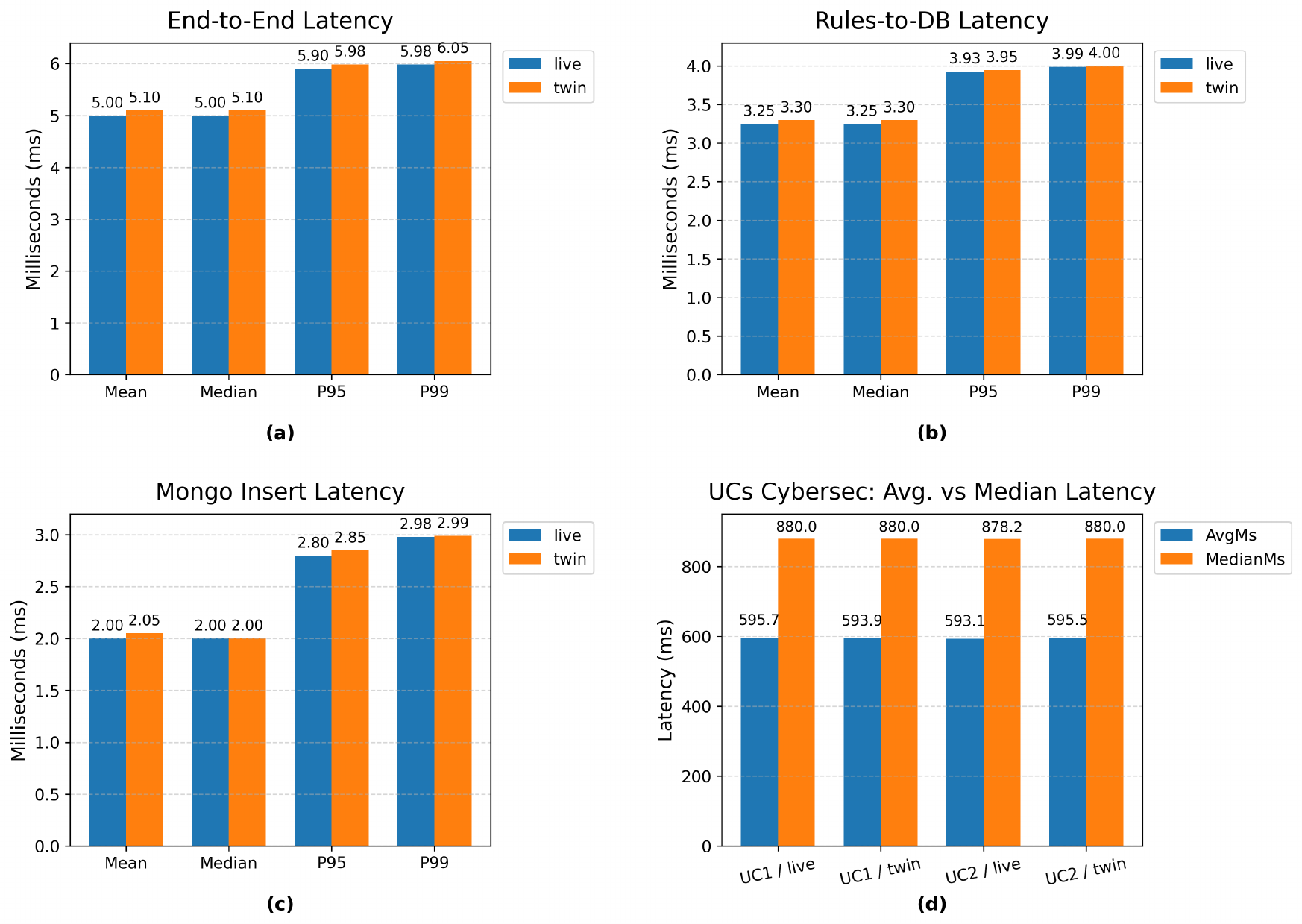}
\caption{Digital twin performance analysis.}
\label{fig:grouped_performance_results}
\end{figure*}

Figure~\ref{fig:hysecdt_detection_analysis}(b) reports detection-time reductions of \textbf{16.52\%} for campaign C0012 and \textbf{21.56\%} for campaign C0025. These gains are achieved without introducing measurable system overhead. Figure~\ref{fig:grouped_performance_results}(a)--(c) shows that end-to-end latency, rules-to-database processing, and MongoDB insert latency remain closely aligned between Live and Twin channels. Under sustained load, Fig.~\ref{fig:grouped_performance_results}(d) further demonstrates stable throughput and consistent average versus median latency behavior across both scenarios.

These results demonstrate that \textbf{combining deterministic rules with fuzzy inference improves detection capability}, particularly for \textbf{ambiguous and multi-signal cyber--physical behaviors}. The outputs align with NIST CSF categories (\textit{DE.DP-5} and \textit{RS.RP-1}), supporting both enhanced detection and effective response execution in operational CPS environments.

\begin{table}[htbp]
\caption{TTP-to-threat mapping for benchmark campaigns}
\label{tab:9-ttp-threat-mapping}
\centering
\begin{tabularx}{\linewidth}{@{}c c X@{}}
\toprule
\textbf{Campaign} & \textbf{TTP} & \textbf{Threat Type} \\
\midrule
C0012 & T1021.002 & Unauthorized remote access to light device via SMB \\
C0012 & T1059 & Malicious command execution affecting device state \\
C0012 & T1112 & Registry/configuration tampering to misreport state \\
C0025 & T0850 & Logical--physical state mismatch: light reported off with low lux \\
C0025 & T0850 & Digital twin state spoofing: light reported on while off \\
\bottomrule
\end{tabularx}
\vspace{-3mm}
\end{table}

\begin{table}[htbp]
\caption{NIST-tagged benchmark results for Durable Rules and Hybrid reasoning engines}
\label{tab:10-ttp-threat-mapping}
\centering
\begin{tabularx}{\linewidth}{@{}l c c c c X@{}}
\toprule
\textbf{Engine} & \textbf{Campaign} & \textbf{TTP} & \textbf{Detection Time (ms)} & \textbf{CSF Tag} & \textbf{Timestamp} \\
\midrule
Durable & C0012 & T0801 & 45.67 & DE.DP-5 & 2025-06-18 14:05:12 \\
Hybrid  & C0012 & T0801 & 38.12 & RS.RP-1 & 2025-06-18 14:05:12 \\
Durable & C0025 & T0850 & 52.34 & DE.DP-5 & 2025-06-18 14:08:27 \\
Hybrid  & C0025 & T0850 & 41.08 & RS.RP-1 & 2025-06-18 14:08:27 \\
\bottomrule
\end{tabularx}
\vspace{-3mm}
\end{table}

\begin{table}[htbp]
\caption{Comparative detection times and NIST CSF alignment for campaign C0012 (TTP: T0801)}
\label{tab:11_nist_ttp_detailed}
\centering
\footnotesize
\setlength{\tabcolsep}{4pt}
\renewcommand{\arraystretch}{1.1}
\begin{tabular}{@{}l c c c c p{4.2cm}@{}}
\toprule
\textbf{Engine} &
\textbf{Campaign} &
\textbf{TTP} &
\textbf{Detection Time (ms)} &
\textbf{CSF Tag} &
\textbf{CSF Description} \\
\midrule
Durable & C0012 & T0801 & 45.67 & DE.DP-5 & Detection processes are continuously improved \\
Hybrid & C0012 & T0801 & 38.12 & RS.RP-1 & Response plan is executed during or after an incident \\
\bottomrule
\end{tabular}
\vspace{-3mm}
\end{table}

\section{Discussion}
\label{sec:7-discussion}
\subsection{Interpretation of Results}

The experimental evaluation shows that HySecTwin can support cybersecurity monitoring in CPS environments while preserving the timing constraints required for stable operation. Across the assessed scenarios, both \textbf{\textit{PT}} and \textbf{\textit{DT}} channels maintained tightly bounded latency, minimal processing divergence, and stable throughput under sustained load. This indicates that semantic reasoning can be integrated into the monitoring pipeline with negligible overhead, while most delay arises from communication and data persistence rather than inference. These findings provide the performance foundation for the broader results discussed below.
\newline 
\textbf{Performance robustness.}
From a systems perspective, the results confirm that the proposed framework maintains predictable and stable performance across both nominal and adversarial conditions. The close alignment between physical and Digital Twin channels indicates that analytics executed within the twin environment can proceed concurrently with CPS operation while preserving temporal consistency. This demonstrates that the framework can provide continuous security visibility without disrupting core control processes.
\newline 
\textbf{Hybrid reasoning effectiveness.}
The evaluation also highlights the benefits of combining deterministic and fuzzy inference mechanisms. Compared with deterministic rules alone, the hybrid reasoning engine achieves faster threat recognition and more stable detection behavior across benchmark campaigns. This suggests that integrating crisp rule logic with graded reasoning improves sensitivity to weak, partial, or evolving indicators of compromise. Such behavior is particularly valuable in CPS environments, where malicious activity may emerge gradually through subtle state deviations rather than explicit rule violations.
\newline 
\textbf{Behavioral fidelity of the Digital Twin.}
The results further indicate that the Digital Twin provides a sufficiently faithful representation of CPS operational behavior to support cybersecurity analytics. Multiple IIoT devices, sensing states, actuation events, and environmental variables were represented through the semantic model, enabling reasoning over interactions rather than isolated signals. The close correspondence between physical and twin performance characteristics suggests that state transitions are mirrored consistently, allowing the twin to function as a trustworthy analytical substrate for monitoring and threat interpretation.
\newline 
\textbf{Cybersecurity analysis of complex CPS.}
The experiments demonstrate that the framework can support higher-level cybersecurity analysis in multi-component CPS environments. By linking semantic system states with ATT\&CK aligned tactics and techniques, the framework enables device observations to be interpreted as structured threat behaviors. This transforms the Digital Twin from a passive replica into an active security analysis platform supporting secure-by-design principles through explainable detection, contextual situational awareness, and informed defensive decision-making.

\subsection{Key Contributions of the HySecTwin Framework}

The evaluation results highlight three primary contributions of the HySecTwin framework: 
(i) a cybersecurity-oriented Digital Twin development that embeds reasoning directly within CPS pipelines, 
(ii) a knowledge-driven semantic modeling layer enabling contextual interpretation of device behavior, and 
(iii) a hybrid reasoning mechanism combining deterministic and fuzzy inference for explainable threat detection.
\newline 
\textbf{1. Digital Twin Development.} A central contribution of this work is the development of a cybersecurity-oriented DT architecture that functions as an analytical monitoring layer rather than a passive system replica. The synchronized interaction between the \textbf{\textit{PT}} and \textbf{\textit{DT}} enables continuous monitoring of CPS telemetry while preserving operational timing constraints. As demonstrated by the latency measurements in Tables~\ref{tab:4-uc_latency_live} and~\ref{tab:5-latency-twin}, the mirrored analytics pipeline introduces negligible overhead, confirming that security reasoning can be embedded directly within DT infrastructures without affecting system responsiveness. This architecture enables real-time anomaly detection and contextual threat analysis while maintaining the operational stability required in industrial CPS environments.
\newline 
\textbf{2. Semantic Modeling.} Another key contribution is the development of a knowledge-driven semantic representation that transforms raw CPS telemetry into machine-interpretable knowledge. The semantic model enables a structured representation of device attributes, operational states, and system relationships using the SAREF ontology. This semantic abstraction provides a consistent vocabulary for modeling CPS components, including sensors, actuators, operational states, and environmental conditions. Constructing semantic models for CPS environments remains a complex task that often requires extensive domain expertise and manual metadata population~\cite{Feng2022}. In addition to annotating CPS attributes, semantic modeling must also capture the relationships between system entities to enable meaningful reasoning over operational states. Prior research has explored ontology-driven approaches to address these challenges. For example, Fang et al.~\cite{ISI:000368652500002} propose a cloud service ontology model incorporating fuzzy extensions within OWL2 to support structured knowledge representation across distributed systems. In the HySecTwin framework, the semantic model was constructed by extending two core components: SAREF Core v3.1.1 based ontology and a domain-specific semantic model described in Section~\ref{sec:3-Background}. These components form the foundation of the semantic model used by the machine reasoning component, enabling structured CPS monitoring and contextual interpretation of device behavior.
\newline 
\textbf{3.Hybrid Reasoning Mechanism.} The third major contribution is the integration of a hybrid reasoning mechanism that combines deterministic rule-based inference with fuzzy reasoning. Deterministic rules provide precise detection of explicit policy violations, such as unauthorized command execution, configuration tampering, or breaches of predefined thresholds. However, purely rule-based reasoning systems are inherently restrictive because they rely on strictly defined logical conditions and may fail to detect ambiguous or gradually emerging anomalies. Hybrid reasoning extends this capability by enabling the aggregation of weak signals across multiple indicators of system behavior. As demonstrated in Tables~\ref{tab:10-ttp-threat-mapping} and~\ref{tab:11_nist_ttp_detailed} and illustrated in Fig.~\ref{fig:hysecdt_detection_analysis}, the hybrid reasoning engine improves detection responsiveness compared with deterministic reasoning while maintaining interpretability. The HySecTwin framework uses a Rete-based reasoning engine, well-suited for constrained IoT and IIoT environments due to its efficient incremental reasoning~\cite{Iorga2018}. Unlike traditional engines with high computational and memory overhead, Rete supports efficient fact management through incremental insertion and deletion. Insertion separates newly observed facts from previously processed ones in alpha memory nodes, while deletion preserves consistency by linking implicit and explicit facts via additional data structures between alpha and beta nodes~\cite{Bento2022}. These features enable scalable reasoning across edge-to-enterprise architectures while maintaining responsiveness in dynamic CPS environments.

\subsection{Comparison with Machine Learning Approaches}

Many contemporary CPS security solutions rely heavily on machine learning and deep learning techniques for anomaly and threat detection. Such approaches can identify complex behavioral patterns and nonlinear dependencies, but they often require large volumes of labeled data, periodic retraining, and careful handling of concept drift as system behavior evolves~\cite{jadidi2022security,xu2023digital}. In safety-critical environments, these models may also operate as opaque decision systems, limiting operator trust, auditability, and forensic interpretation.
In contrast, the HySecTwin framework adopts a reasoning-driven approach grounded in semantic knowledge representation, deterministic rules, and fuzzy inference. This approach offers several advantages for critical infrastructure environments. First, reasoning outputs remain interpretable and traceable, allowing operators to understand why anomalies are detected and which system states triggered a response. Second, the framework does not depend on extensive training datasets that are often unavailable or costly to curate in operational CPS deployments. Third, explicit rules enable direct encoding of engineering constraints, safety policies, and security logic, while fuzzy reasoning provides robustness to uncertainty, noisy telemetry, and partial evidence.
Recent studies have also recognized the need to move beyond purely black-box detection models. Digital twin-assisted anomaly detection methods improve contextual awareness through system replicas, while newer hybrid architectures combine learned models with structured reasoning to improve explainability and operational trustworthiness~\cite{xu2023digital,kampourakis2026systematic}. HySecTwin aligns with this direction but emphasizes lightweight semantic reasoning and policy-driven detection suitable for deterministic CPS environments. Rather than replacing machine learning methods, the proposed architecture should be viewed as complementary. Data-driven models may remain effective for discovering previously unseen patterns, whereas the HySecTwin reasoning layer can provide transparent decision support, policy compliance checking, and contextual interpretation of CPS telemetry within a structured knowledge framework.

\subsection{Limitations}

While the proposed Digital Twin-based cybersecurity framework can monitor cyber-physical system behavior and reason about anomalies and hidden threats, it has some limitations. Its effectiveness depends on the fidelity of the virtual representation and the availability of accurate telemetry from the physical system, which this work achieves. However, minor discrepancies between the physical and digital systems may affect anomaly interpretation, particularly in heterogeneous industrial environments with diverse devices and protocols \cite{Varghese2022}. Moreover, the reasoning layer relies on semantic rules and behavioral relationships derived from the system model. Although this supports interpretable and explainable decisions, detecting entirely novel attack strategies may require complementary data-driven approaches. Machine learning and deep learning methods can detect complex or unseen anomalies, but they often lack explainability and require large datasets and significant computational resources \cite{sayghe25,10431655}. In contrast, rule-based semantic approaches are more transparent, yet they depend on complete and accurate system knowledge \cite{ahmad-bridgeICS2025-kGs,10431655}.

\section{Conclusion and Future Work}
\label{sec:8-Conclusion}
This paper presents HySecTwin, a knowledge-driven DT framework designed to strengthen cybersecurity monitoring and operational resilience in a CPS. The proposed architecture integrates real-time DT synchronization with structured system knowledge and a hybrid reasoning engine, combining deterministic rule-based inference and fuzzy reasoning. This combination enables the framework to analyse CPS behavior beyond rigid rule violations by capturing subtle anomalies and contextual relationships within system operations. As a result, HySecTwin supports explainable and context-aware threat detection, addressing a key limitation of black-box cybersecurity solutions. Evaluation using a representative CPS testbed and MITRE ATT\&CK for ICS–inspired attack scenarios confirms the feasibility of integrating knowledge-driven Digital Twins with hybrid reasoning for real-time security monitoring. Future research will extend this framework by incorporating Agentic AI capabilities, enabling advanced knowledge interpretation, automated reasoning support, and adaptive cybersecurity analytics. Integrating Human-in-the-Loop collaboration with these intelligent agents will further support supervised decision-making, explainable threat analysis, and coordinated defense strategies across large-scale cyber–physical infrastructures.

\section*{Declarations}

\subsection*{Data Availability.} The data and materials supporting the findings of this research, including datasets, source code, experimental scripts, and Ontology Models, are publicly available at the \href{https://github.com/ahmadspm/HybridSec-Digital-Twin-CSP-Modelling/tree/main}{Hybrid-Security Twin Project}.

 \subsection*{Funding.}
This work was supported by funding from the Cyber Security Cooperative Research Center (CSCRC) Australia through its PhD Scholarship program.. 

\subsection*{Acknowledgments.} This project was primarily supported by the Cyber Security Cooperative Research Center (CSCRC), Australia, and the School of Science, Edith Cowan University, Australia. The authors gratefully acknowledge the resources provided for testbed establishment, experimentation, and Digital Twin development.

\section*{CRediT authorship contribution statement}
Writing – original draft: D.H, A.M; Writing – review \& editing: A.M, S.N, I.H.S, L.S, H.J; Corresponding Author: A.M

\bibliography{references}

@misc{oasis-mqtt-v3_1_1,
  title        = {MQTT Version 3.1.1},
  author       = {{OASIS}},
  howpublished = {OASIS Standard},
  year         = {2014},
  month        = {10},
  url          = {https://docs.oasis-open.org/mqtt/mqtt/v3.1.1/os/mqtt-v3.1.1-os.html},
  note         = {Accessed 2025-08-24}
}

@misc{oasis-mqtt-v5_0,
  title        = {MQTT Version 5.0},
  author       = {{OASIS}},
  howpublished = {OASIS Standard},
  year         = {2019},
  month        = {03},
  url          = {https://docs.oasis-open.org/mqtt/mqtt/v5.0/os/mqtt-v5.0-os.html},
  note         = {Accessed 2025-08-24}
}

@techreport{nist-sp-800-82r2,
  title        = {Guide to Industrial Control Systems (ICS) Security},
  author       = {Stouffer, Keith and Falco, Joseph and Scarfone, Karen},
  institution  = {National Institute of Standards and Technology (NIST)},
  type         = {Special Publication},
  number       = {800-82, Revision 2},
  year         = {2015},
  doi          = {10.6028/NIST.SP.800-82r2},
  url          = {https://doi.org/10.6028/NIST.SP.800-82r2},
  note         = {Accessed 2025-08-24}
}

@techreport{nist-sp-800-55r2,
  title        = {Performance Measurement Guide for Information Security},
  author       = {{National Institute of Standards and Technology (NIST)}},
  institution  = {NIST},
  type         = {Special Publication},
  number       = {800-55, Revision 2},
  year         = {2023},
  url          = {https://csrc.nist.gov/publications/detail/sp/800-55/rev-2/final},
  note         = {Accessed 2025-08-24}
}

@inproceedings{cooper2010ycsb,
author = {Cooper, Brian F. and Silberstein, Adam and Tam, Erwin and Ramakrishnan, Raghu and Sears, Russell},
title = {Benchmarking cloud serving systems with YCSB},
year = {2010},
isbn = {9781450300360},
publisher = {Association for Computing Machinery},
address = {New York, NY, USA},
doi = {10.1145/1807128.1807152},
booktitle = {Proceedings of the 1st ACM Symposium on Cloud Computing},
pages = {143–154},
numpages = {12},
location = {Indianapolis, Indiana, USA},
series = {SoCC '10}
}

@article{rashid2022hybridids,
  author    = {Rashid, M. M. and Alazab, M. and Anwar, A. and Khan, S.},
  title     = {A hybrid machine learning and fuzzy logic-based intrusion detection system for industrial control systems},
  journal   = {Computers \& Security},
  volume    = {114},
  year      = {2022},
  pages     = {102578},
  doi       = {10.1016/j.cose.2022.102578}
}

@book{klir1995fuzzy,
  author    = {Klir, George J. and Yuan, Bo},
  title     = {Fuzzy Sets and Fuzzy Logic: Theory and Applications},
  year      = {1995},
  publisher = {Prentice Hall},
  address   = {Upper Saddle River, NJ, USA}
}

@article{wang2021faultdiag,
  author    = {Wang, Jian and Zhou, Hua and Li, Peng},
  title     = {Rule-based and fuzzy logic hybrid reasoning for intelligent fault diagnosis in complex systems},
  journal   = {Expert Systems with Applications},
  volume    = {182},
  year      = {2021},
  pages     = {115220},
  doi       = {10.1016/j.eswa.2021.115220}
}

@article{Alnowaiser2023,
author = {Alnowaiser, Kholood K. and Ahmed, Moataz A.},
doi = {10.1007/s13369-022-07459-0},
issn = {21914281},
journal = {Arabian Journal for Science and Engineering},
number = {2},
pages = {1075--1095},
publisher = {Springer Berlin Heidelberg},
title = {{Digital Twin: Current Research Trends and Future Directions}},
url = {https://doi.org/10.1007/s13369-022-07459-0},
volume = {48},
year = {2023}
}

@article{Empl2022,
author = {Empl, Philip and Schlette, Daniel and Zupfer, Daniel and Pernul, G{\"{u}}nther},
doi = {10.1145/3538969.3538975},
isbn = {9781450396707},
journal = {ACM International Conference Proceeding Series},
pages = {1--10},
title = {{SOAR4IoT: Securing IoT Assets with Digital Twins}},
year = {2022}
}

@inproceedings{Kharlamov2019,
  author={Kharlamov, Evgeny and Martin-Recuerda, Francisco and Perry, Brandon and Cameron, David and Fjellheim, Roar and Waaler, Arild},
  booktitle={2018 IEEE International Conference on Big Data (Big Data)}, 
  title={Towards Semantically Enhanced Digital Twins}, 
  year={2018},
  volume={},
  number={},
  pages={4189-4193},
  doi={10.1109/BigData.2018.8622503}}

@article{Varghese2022,
  author={Varghese, Seba Anna and Dehlaghi Ghadim, Alireza and Balador, Ali and Alimadadi, Zahra and Papadimitratos, Panos},
  booktitle={2022 IEEE International Conference on Pervasive Computing and Communications Workshops and other Affiliated Events (PerCom Workshops)}, 
  title={Digital Twin-based Intrusion Detection for Industrial Control Systems}, 
  year={2022},
   pages={611-617},
  doi={10.1109/PerComWorkshops53856.2022.9767492}}

@InProceedings{Bento2022,
author="Bento, Alexandre
and M{\'e}dini, Lionel
and Singh, Kamal
and Laforest, Fr{\'e}d{\'e}rique",
title="Do Arduinos Dream of Efficient Reasoners?",
booktitle="The Semantic Web",
year="2022",
publisher="Springer International Publishing",
address="Cham",
pages="289--304",
isbn="978-3-031-06981-9"
}

@INPROCEEDINGS{Balta2019,
  author={Balta, Efe C. and Tilbury, Dawn M. and Barton, Kira},
  booktitle={2019 IEEE 15th International Conference on Automation Science and Engineering (CASE)}, 
  title={A Digital Twin Framework for Performance Monitoring and Anomaly Detection in Fused Deposition Modeling}, 
  year={2019},
  pages={823-829},
  doi={10.1109/COASE.2019.8843166},
  ISSN={2161-8089},
  month={Aug},
}

@article{Balta2023,
author = {Balta, Efe C. and Pease, Michael and Moyne, James and Barton, Kira and Tilbury, Dawn M.},
doi = {10.1109/TASE.2023.3243147},
issn = {15583783},
journal = {IEEE Transactions on Automation Science and Engineering},
pages = {1--18},
publisher = {IEEE},
title = {{Digital Twin-Based Cyber-Attack Detection Framework for Cyber-Physical Manufacturing Systems}},
volume = {PP},
year = {2023}
}

@article{Masi2023,
author = {Masi, Massimiliano and Sellitto, Giovanni Paolo and Aranha, Helder and Pavleska, Tanja},
doi = {10.1007/s10270-022-01075-0},
issn = {16191374},
journal = {Software and Systems Modeling},
number = {2},
pages = {689--707},
publisher = {Springer},
title = {{Securing critical infrastructures with a cybersecurity digital twin}},
volume = {22},
year = {2023}
}

@article{Eckhart2019,
author = {Eckhart, Matthias and Ekelhart, Andreas and Weippl, Edgar},
doi = {10.1109/ETFA.2019.8869197},
isbn = {9781728103037},
issn = {19460759},
journal = {IEEE International Conference on Emerging Technologies and Factory Automation, ETFA},
pages = {1222--1225},
publisher = {IEEE},
title = {{Enhancing Cyber Situational Awareness for Cyber-Physical Systems through Digital Twins}},
volume = {2019-September},
year = {2019}
}

@article{Eckhart2023,
author = {Eckhart, Matthias and Ekelhart, Andreas and Allison, David and Almgren, Magnus and Ceesay-Seitz, Katharina and Janicke, Helge and Nadjm-Tehrani, Simin and Rashid, Awais and Yampolskiy, Mark},
doi = {10.1109/MSEC.2023.3271225},
issn = {15584046},
journal = {IEEE Security and Privacy},
pages = {1--12},
title = {{Security-Enhancing Digital Twins: Characteristics, Indicators, and Future Perspectives}},
year = {2023}
}

@article{Feng2022,
author = {Feng, Zai Wen and Xu, Jia Kang and Mayer, Wolfgang and Huang, Wang Yu and He, Ke Qing and Stumptner, Markus and Grossmann, Georg and Zhang, Hong Yu and Ling, Lin},
doi = {10.1109/HPCC-DSS-SmartCity-DependSys53884.2021.00304},
isbn = {9781665494571},
journal = {2021 IEEE 23rd International Conference on High Performance Computing and Communications, 7th International Conference on Data Science and Systems, 19th International Conference on Smart City and 7th International Conference on Dependability in Sensor, Cl},
pages = {2034--2041},
publisher = {IEEE},
title = {{Automatic Semantic Modeling for Structural Data Source with the Prior Knowledge From Knowledge Graph}},
year = {2022}
}

@article{Iorga2018,
author = {Iorga, Michaela and Feldman, Larry and Barton, Robert and Martin, Michael J},
number = {March},
title = {{NIST Special Publication 500-325 Recommendations of the National Institute of Standards and Technology}},
year = {2018}
}

@article{ISI:000368652500002,
address = {RADARWEG 29, 1043 NX AMSTERDAM, NETHERLANDS},
author = {Fang, Daren and Liu, Xiaodong and Romdhani, Imed and Jamshidi, Pooyan and Pahl, Claus},
doi = {10.1016/j.future.2015.09.025},
issn = {0167739X},
journal = {Future Generation Computer Systems},
month = {mar},
pages = {11--26},
publisher = {Elsevier},
title = {{An agility-oriented and fuzziness-embedded semantic model for collaborative cloud service search, retrieval and recommendation}},
type = {Article},
url = {https://linkinghub.elsevier.com/retrieve/pii/S0167739X15003052},
volume = {56},
year = {2016}
}

@article{Holmesa,
author = {Holmes, David and Papathanasaki, Maria and Maglaras, Leandros and Ferrag, Mohamed Amine and Nepal, Surya and Janicke, Helge},
doi = {10.1109/SEEDA-CECNSM53056.2021.9566277},
isbn = {9781665427425},
journal = {6th South-East Europe Design Automation, Computer Engineering, Computer Networks and Social Media Conference, SEEDA-CECNSM 2021},
title = {{Digital Twins and Cyber Security - solution or challenge?}},
year = {2021}
}

@article{Gerodimos2023,
author = {Gerodimos, Apostolos and Maglaras, Leandros and Ferrag, Mohamed Amine and Ayres, Nick and Kantzavelou, Ioanna},
doi = {10.1016/j.iotcps.2022.12.003},
issn = {26673452},
journal = {Internet of Things and Cyber-Physical Systems},
number = {December 2022},
pages = {1--13},
publisher = {The Authors},
title = {{IoT: Communication protocols and security threats}},
volume = {3},
year = {2023}
}

@book{Eckhart,
author = {Eckhart, Matthias and Ekelhart, Andreas},
booktitle = {Security and Quality in Cyber-Physical Systems Engineering},
doi = {10.1007/978-3-030-25312-7_14},
pages = {383--412},
publisher = {Springer},
address = {Cham},
title = {{Digital Twins for Cyber-Physical Systems Security: State of the Art and Outlook}},
year = {2019}
}

@inproceedings{Bromander201674,
author = {Bromander, S and J{\o}sang, A and Eian, M},
booktitle = {CEUR Workshop Proceedings},
editor = {{Oltramari A. Emmons I.}, Costa P C G Laskey K B},
issn = {16130073},
pages = {74--78},
publisher = {Aachen University},
address = {Aachen},
title = {{Semantic cyberthreat modelling}},
volume = {1788},
year = {2016}
}

@article{Datta201714,
author = {Datta, S P A},
doi = {10.24840/2183-0606_005.003_0003},
issn = {21830606},
journal = {Journal of Innovation Management},
number = {3},
pages = {14--33},
publisher = {Universidade do Porto - Faculdade de Engenharia},
title = {{Emergence of Digital Twins - Is this the March of reason?}},
volume = {5},
year = {2017}
}

@article{Eckhart2018,
author = {Eckhart, Matthias and Ekelhart, Andreas},
doi = {10.1145/3264888.3264892},
isbn = {9781450359924},
issn = {15437221},
journal = {Proceedings of the ACM Conference on Computer and Communications Security},
pages = {36--47},
title = {A specification-based state replication approach for digital twins},
year = {2018}
}

@inproceedings{Eckhartr201861,
author = {Eckhart, Matthias and Ekelhart, Andreas},
title = {Towards Security-Aware Virtual Environments for Digital Twins},
year = {2018},
isbn = {9781450357555},
publisher = {Association for Computing Machinery},
address = {New York, NY, USA},
doi = {10.1145/3198458.3198464},
booktitle = {Proceedings of the 4th ACM Workshop on Cyber-Physical System Security},
pages = {61–72},
numpages = {12},
location = {Incheon, Republic of Korea},
series = {CPSS '18}
}

@article{Gaikwad2020,
author = {Aniruddha Gaikwad and Reza Yavari and Mohammad Montazeri and Kevin Cole and Linkan Bian and Prahalada Rao},
title = {Toward the digital twin of additive manufacturing: Integrating thermal simulations, sensing, and analytics to detect process faults},
journal = {IISE Transactions},
volume = {52},
number = {11},
pages = {1204-1217},
year  = {2020},
publisher = {Taylor \& Francis},
doi = {10.1080/24725854.2019.1701753}
}

@article{Grieves2016,
author = {Grieves, Michael and Vickers, John},
doi = {10.1007/978-3-319-38756-7_4},
isbn = {9783319387567},
journal = {Transdiscipl. Perspect. Complex Syst. New Find. Approaches},
number = {August},
pages = {85--113},
title = {{Digital twin: Mitigating unpredictable, undesirable emergent behavior in complex systems}},
year = {2016}
}

@inproceedings{Kummerow2019,
author = {Kummerow, André and Rösch, Dennis and Monsalve, Cristian and Nicolai, Steffen and Bretschneider, Peter and Brosinsky, Christoph and Westermann, Dirk},
year = {2019},
month = {06},
pages = {1-6},
title = {Challenges and opportunities for phasor data based event detection in transmission control centers under cyber security constraints},
doi = {10.1109/PTC.2019.8810711}
}

@article{Kritzinger2018c,
author = {Kritzinger, Werner and Karner, Matthias and Traar, Georg and Henjes, Jan and Sihn, Wilfried},
doi = {10.1016/j.ifacol.2018.08.474},
issn = {24058963},
journal = {IFAC-PapersOnLine},
number = {11},
pages = {1016--1022},
publisher = {Elsevier B.V.},
title = {{Digital Twin in manufacturing: A categorical literature review and classification}},
url = {https://doi.org/10.1016/j.ifacol.2018.08.474},
volume = {51},
year = {2018}
}

@article{Moser2023,
author = {Moser, Bryan R. and Grossmann, William},
doi = {10.1007/978-3-031-21343-4_24},
isbn = {9783031213434},
journal = {The Digital Twin},
pages = {677--702},
title = {{Digital Twins of Complex Projects}},
year = {2023}
}

@article{Talkhestani2018,
author = {Talkhestani, Behrang Ashtari and Jazdi, Nasser and Schloegl, Wolfgang and Weyrich, Michael},
doi = {10.1016/j.procir.2018.03.166},
isbn = {4971168567},
issn = {22128271},
journal = {Procedia CIRP},
number = {March},
pages = {159--164},
publisher = {Elsevier B.V.},
title = {{Consistency check to synchronize the Digital Twin of manufacturing automation based on anchor points}},
url = {https://doi.org/10.1016/j.procir.2018.03.166},
volume = {72},
year = {2018}
}

@incollection{Sikos2019,
  author = {Sikos, L. F.},
  title = {{OWL} Ontologies in Cybersecurity: Conceptual Modeling of Cyber-Knowledge},
  editor = {Sikos, L. F.},
  booktitle = {{AI in Cybersecurity}},
  publisher = {Springer},
  address = {Cham},
  year = {2019},
  pages = {1--17},
  doi = {10.1007/978-3-319-98842-9_1}
}

@incollection{Sikos2020,
  author = {Leslie F. Sikos},
  title = {The Formal Representation of Cyberthreats for Automated Reasoning},
  editor = {Leslie F. Sikos and Kim-Kwang Raymond Choo},
  booktitle = {{Data Science in Cybersecurity and Cyberthreat Intelligence}},
  publisher = {Springer},
  address = {Cham},
  year = {2020},
  pages = {1--12},
  doi = {10.1007/978-3-030-38788-4_1}
}

@inproceedings{Horrocksetal2006,
  title={The even more irresistible SROIQ.},
  author={Horrocks, Ian and Kutz, Oliver and Sattler, Ulrike},
  journal={Kr},
  volume={6},
  pages={57--67},
  year={2006}
}

@inbook{Sikos2015,
  author = {Leslie F. Sikos}, 
  title = {{Mastering Structured Data on the Semantic Web}},
  pages = {37--38},
  publisher = {Apress},
  address = {Berkeley},
  year = {2015},
  doi = {10.1007/978-1-4842-1049-9}
}

@inproceedings{Muralidharanetal2020,
  author={Muralidharan, Shapna and Yoo, Byounghyun and Ko, Heedong},
  booktitle={2020 IEEE International Conference on Consumer Electronics (ICCE)}, 
  title={Designing a Semantic Digital Twin model for IoT}, 
  year={2020},
  volume={},
  number={},
  pages={1-2},
  doi={10.1109/ICCE46568.2020.9043088}}

@misc{EclpliseDITTO,
  title = {Eclipse. 2024. Eclipse Ditto},
  howpublished = {\url{http://https://github.com/eclipse/ditto}},
  note = {Accessed: 10-06-2024}
}

@misc{AmazonTwinmaker,
  title = {Amazon. 2024. AWS IoT TwinMaker
},
  howpublished = {\url{https://aws.amazon.com/iot-twinmaker/}},
  note = {Accessed: 05-02-2024}
}

@misc{AzureDigitalTwin,
  title = {Microsoft. 2024. Azure Digital Twin},

  howpublished = {\url{https://docs.microsoft.com/en-gb/azure/digital-twins}},
  note = {Accessed: 02-10-2024}
}

@INPROCEEDINGS {10431655,
author = { Mohsin, Ahmad and Janicke, Helge and Nepal, Surya and Holmes, David },
booktitle = { 2023 5th IEEE International Conference on Trust, Privacy and Security in Intelligent Systems and Applications (TPS-ISA) },
title = {{ Digital Twins and the Future of Their Use Enabling Shift Left and Shift Right Cybersecurity Operations }},
year = {2023},
volume = {},
ISSN = {},
pages = {277-286},
doi = {10.1109/TPS-ISA58951.2023.00042},
publisher = {IEEE Computer Society},
address = {Los Alamitos, CA, USA},
month =Nov}

@article{SARKER2024935,
title = {Explainable AI for cybersecurity automation, intelligence and trustworthiness in digital twin: Methods, taxonomy, challenges and prospects},
journal = {ICT Express},
volume = {10},
number = {4},
pages = {935-958},
year = {2024},
issn = {2405-9595},
doi = {https://doi.org/10.1016/j.icte.2024.05.007},
author = {Iqbal H. Sarker and Helge Janicke and Ahmad Mohsin and Asif Gill and Leandros Maglaras}
}

@inproceedings{liu2016rule,
  author={Liu, Han and Gegov, Alexander},
  booktitle={2016 IEEE 8th International Conference on Intelligent Systems (IS)}, 
  title={Rule based systems and networks: Deterministic and fuzzy approaches}, 
  year={2016},
   pages={316-321},
  doi={10.1109/IS.2016.7737440}}

@inproceedings{anicic2010rule,
  title={A rule-based language for complex event processing and reasoning},
  author={Anicic, Darko and Fodor, Paul and Rudolph, Sebastian and St{\"u}hmer, Roland and Stojanovic, Nenad and Studer, Rudi},
  booktitle={International Conference on Web Reasoning and Rule Systems},
  pages={42--57},
  year={2010},
  organization={Springer}
}

@article{Empl2023,
  author       = {Empl, Philip and Pernul, G{\"u}nther},
  title        = {Digital-Twin-Based Security Analytics for the Internet of Things},
  journal      = {Information},
  volume       = {14},
  number       = {2},
  pages        = {95},
  year         = {2023},
  publisher    = {MDPI},
  doi          = {10.3390/info14020095},
  url          = {https://www.mdpi.com/2078-2489/14/2/95}
}

@inproceedings{Sikos2018,
  author = {Leslie F. Sikos},
  title = {Handling Uncertainty and Vagueness in Network Knowledge Representation for Cyberthreat Intelligence},
  booktitle = {{2018 IEEE International Conference on Fuzzy Systems}},
  publisher = {{IEEE}},
  address = {New York},
  year  = {2018},
  doi = {10.1109/FUZZ-IEEE.2018.8491686}
}

@Article{sayghe25,
AUTHOR = {Sayghe, Ali},
TITLE = {Digital Twin-Driven Intrusion Detection for Industrial SCADA: A Cyber-Physical Case Study},
JOURNAL = {Sensors},
VOLUME = {25},
YEAR = {2025},
NUMBER = {16},
ARTICLE-NUMBER = {4963},
PubMedID = {40871827},
ISSN = {1424-8220},
DOI = {10.3390/s25164963}
}

@misc{ahmad-bridgeICS2025-kGs,
      title={BRIDG-ICS: AI-Grounded Knowledge Graphs for Intelligent Threat Analytics in Industry~5.0 Cyber-Physical Systems}, 
      author={Padmeswari Nandiya and Ahmad Mohsin and Ahmed Ibrahim and Iqbal H. Sarker and Helge Janicke},
      year={2026},
      eprint={2512.12112},
      archivePrefix={arXiv},
      primaryClass={cs.CR},
      url={https://arxiv.org/abs/2512.12112}, 
}

@inproceedings{jadidi2022security,
  author={Jadidi, Zahra and Pal, Shantanu and K, Nithesh Nayak and Selvakkumar, Arawinkumaar and Chang, Chih-Chia and Beheshti, Maedeh and Jolfaei, Alireza},
  booktitle={2022 International Conference on Computer Communications and Networks (ICCCN)}, 
  title={Security of Machine Learning-Based Anomaly Detection in Cyber Physical Systems}, 
  year={2022},
  volume={},
  number={},
  pages={1-7},
  doi={10.1109/ICCCN54977.2022.9868845}}

@article{xu2023digital,
  author  = {Qianwen Xu and Shaukat Ali and Tao Yue},
  title   = {Digital Twin-Based Anomaly Detection with Curriculum Learning in Cyber-Physical Systems},
  journal = {ACM Transactions on Software Engineering and Methodology},
  volume  = {32},
  number  = {5},
  pages   = {122:1--122:31},
  year    = {2023},
  publisher = {ACM},
  doi     = {10.1145/3597507}
}

@article{kampourakis2026systematic,
  author  = {Konstantinos E. Kampourakis and Vasileios Gkioulos and
             Stefanos Katsikas},
  title   = {Systematic Integration of Digital Twins and Constrained LLMs for Interpretable Cyber-Physical Anomaly Detection},
  journal = {arXiv preprint arXiv:2604.03790},
  year    = {2026},
  eprint  = {2604.03790},
  archivePrefix = {arXiv},
  primaryClass = {cs.CR},
  url      = {https://arxiv.org/abs/2604.03790}
}

@techreport{nistcsf20,
  author      = {{National Institute of Standards and Technology (NIST)}},
  title       = {The {NIST} Cybersecurity Framework ({CSF}) 2.0},
  institution = {National Institute of Standards and Technology},
  type        = {NIST Cybersecurity White Paper},
  number      = {29},
  address     = {Gaithersburg, MD, USA},
  year        = {2024},
  url         = {https://nvlpubs.nist.gov/nistpubs/CSWP/NIST.CSWP.29.pdf}
}

\end{document}